\documentclass[10pt,twocolumn,letterpaper]{article}

\usepackage{iccv}
\usepackage{times}
\usepackage{epsfig}
\usepackage{graphicx}
\usepackage{amsmath}
\usepackage{amssymb}

\usepackage{comment}
\usepackage{enumitem}
\usepackage{multirow}

\ifdefined\reviewversion
\usepackage[pagebackref=true,breaklinks=true,letterpaper=true,colorlinks,bookmarks=false]{hyperref}
\else
\usepackage[breaklinks=true,colorlinks,bookmarks=false]{hyperref}
\iccvfinalcopy
\fi

\newcommand{\norm}[1]{\left\lVert#1\right\rVert}
\def\iccvPaperID{6826} %

\ificcvfinal\fi

\begin{document}

\title{Continuous-Time Spatiotemporal Calibration of \\
a Rolling Shutter Camera---IMU System}

\author{Jianzhu Huai \\
Wuhan University\\
Wuhan Hubei China\\
{\tt\small jianzhu.huai@whu.edu.cn}
\and
Yuan Zhuang\thanks{Corresponding author} \quad\quad Qicheng Yuan \\
Wuhan University\\
Wuhan Hubei China\\
{\tt\small yuan.zhuang@whu.edu.cn}
\and Yukai Lin\\
Huawei Inc.\\
Shanghai China\\
{\tt\small linyukai@outlook.com}
}

\maketitle
\ificcvfinal\thispagestyle{empty}\fi

\begin{abstract}
The rolling shutter (RS) mechanism is widely used by consumer-grade cameras,
which are essential parts in smartphones and autonomous vehicles.
The RS effect leads to image distortion upon relative motion between a camera and the scene.
This effect needs to be considered in video stabilization, 
structure from motion, and vision-aided odometry, for which recent studies have improved 
earlier global shutter (GS) methods by accounting for the RS effect.
However, it is still unclear how the RS affects spatiotemporal calibration of the camera in a sensor assembly, 
which is crucial to good performance in aforementioned applications.

This work takes the camera-IMU system as an example and looks into the RS effect on its spatiotemporal calibration.
To this end, we develop a calibration method for a RS-camera-IMU system with continuous-time B-splines by using a calibration target.
Unlike in calibrating GS cameras, every observation of a landmark on the target
has a unique camera pose fitted by continuous-time B-splines.
With simulated data generated from four sets of public calibration data,
we show that RS can noticeably affect the extrinsic parameters,
causing errors about 1$^\circ$ in orientation and 2 cm in translation with 
a RS setting as in common smartphone cameras.
With real data collected by two industrial camera-IMU systems,
we find that considering the RS effect gives more accurate and consistent spatiotemporal calibration.
Moreover, our method also accurately calibrates the inter-line delay of the RS.
The code for simulation and calibration is publicly available\footnote{\url{https://github.com/JzHuai0108/kalibr}}.
\end{abstract}

\section{Introduction}
Consumer-grade cameras usually have a rolling shutter (RS) mechanism which 
causes consecutive rows of an image to be captured with an inter-line delay.
RS cameras are widely used in smartphones \cite{huaiMobileARSensor2019},
mixed reality products \cite{kerlDense2015},
and autonomous driving cars \cite{sunScalability2020}.
Thanks to the RS feature, these cameras have been used to measure distances \cite{kimObject2020}, 
to estimate motion at a high frequency \cite{bapatRolling2018},
and to identify modulated flickering LEDs \cite{zhuangSurvey2018}.
On the other hand, RS introduces image distortion when there is relative motion
between the camera and the scene.
For better performance, this distortion needs to be considered in applications sensitive to motion.
In response, methods tailored to RS cameras have been proposed for video stablization 
\cite{ringabyEfficient2012}, camera calibration \cite{othRolling2013},
structure from motion \cite{imAccurate2019},
vision-aided odometry \cite{liVisionaidedInertialNavigation2014},
dense mapping \cite{kerlDense2015}, \etc.

For applications like vision-aided odometry and dense mapping, a camera is usually rigidly attached to other sensors,
\eg, depth cameras, lidars, IMUs, to provide complementary data.
One important step of these applications is the spatiotemporal calibration of these sensors.
Existing calibration methods usually assume that the camera uses a global shutter (GS), \eg, \cite{rehderExtending2016}.
For specific sensor assemblies, \eg, a lidar-camera rig, a RGB-D sensor, it is possible 
to estimate the extrinsic parameters using standstill data, \eg,
\cite{kangAutomatic2020}.
But for a lidar-camera rig, many recent algorithms require motion between the sensor system and the scene
for fast calibration \cite{parkSpatiotemporal2020,nowickiSpatiotemporal2020} and 
hence these methods were only validated with GS cameras.
For camera-IMU systems, the extrinsic calibration always requires egomotion where the RS skew comes into play.
But there has been limited spatiotemporal calibration methods for RS cameras.
Consequently, it has been unclear how the RS affects the extrinsic calibration.

Taking the camera-IMU system as an example, this paper looks into the effect of RS on spatiotemporal calibration.
We formulate the calibration problem with continuous-time B-splines \cite{furgaleContinuoustime2012} which 
allow interpolating a unique camera pose for each observation in a RS image, 
precisely handling the RS effect.
By differentiating the B-splines, it is straightforward to accommodate the IMU data.
This formulation has been shown to improve simultaneous localization and mapping
(SLAM) with a RGB-D camera \cite{kerlDense2015}, 
and spatiotemporal calibration of combinations of a GS camera, an IMU, 
and a laser range finder \cite{rehderGeneralApproachSpatiotemporal2016}.
In this regard, our work extends the existing calibration methods to deal with the RS effect.

The proposed method was evaluated with simulated data generated from four sets of public calibration data,
and with real data captured by two industrial camera-IMU systems.
The simulation and real data tests showed that considering the RS effect in spatiotemporal calibration often
improved relative orientation by 1$^\circ$ and relative location by 2 cm from results of a GS-based calibration approach.

The contributions of this paper are as follows.
\setlist{nolistsep}
\begin{enumerate}[noitemsep]
\item We propose a continuous-time B-spline-based approach for spatiotemporal calibration of systems composed of a RS camera and an IMU.
Our novelty lies in considering the RS effect in continuous-time spatiotemporal calibration (Section~\ref{sec:method}).
\item We quantify the RS effect on extrinsic and temporal calibration with simulation and real data,
answering questions about the necessity of accounting for the RS effect (Section~\ref{sec:experiments}).
\item As a byproduct, the proposed approach also provides accurate estimates for the line delay (Section~\ref{subsubsec:bluefox}).
\end{enumerate}

\begin{figure}[t]
\centering
\includegraphics[width=0.7\linewidth]{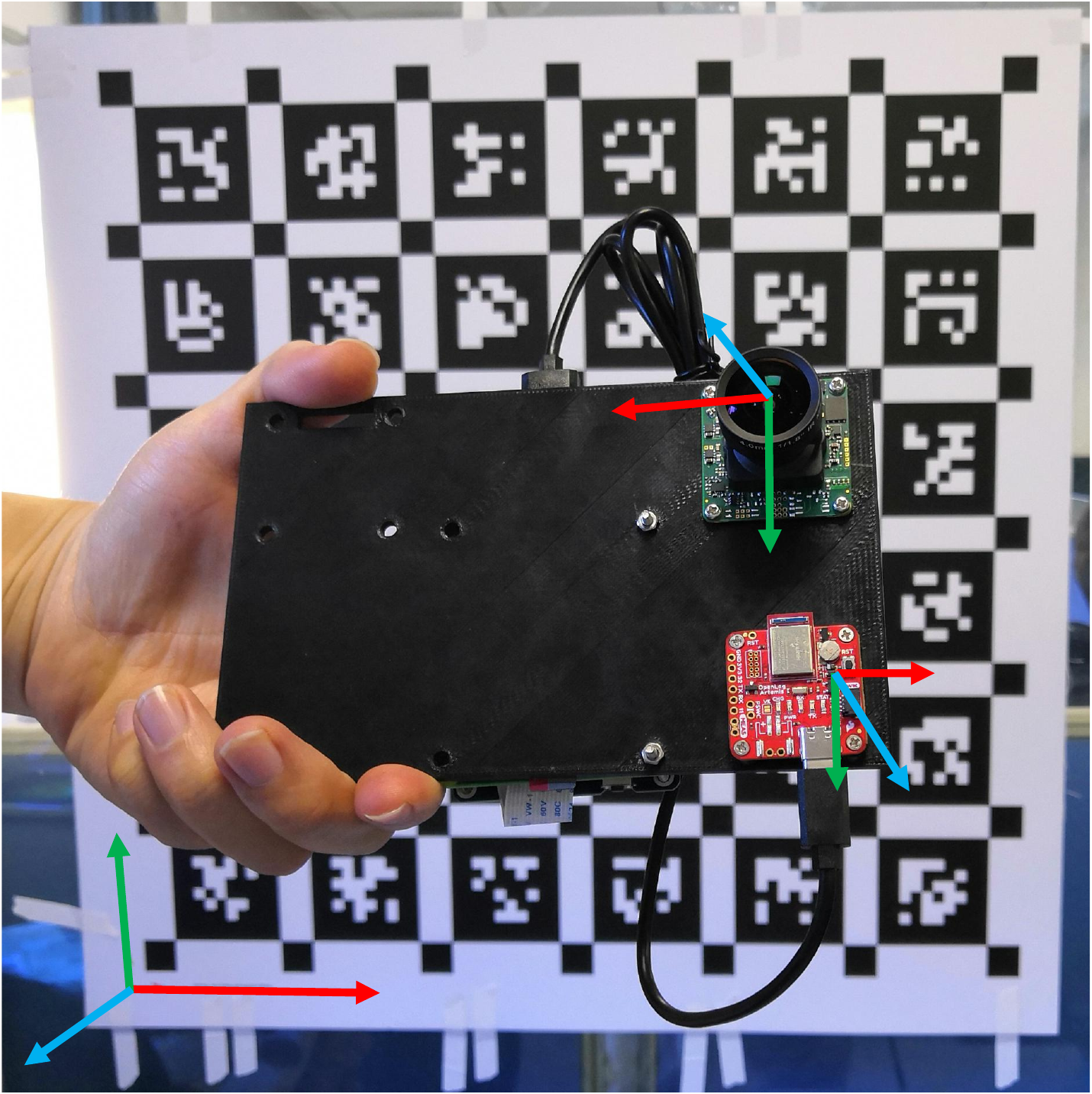} \\
\includegraphics[width=0.7\linewidth]{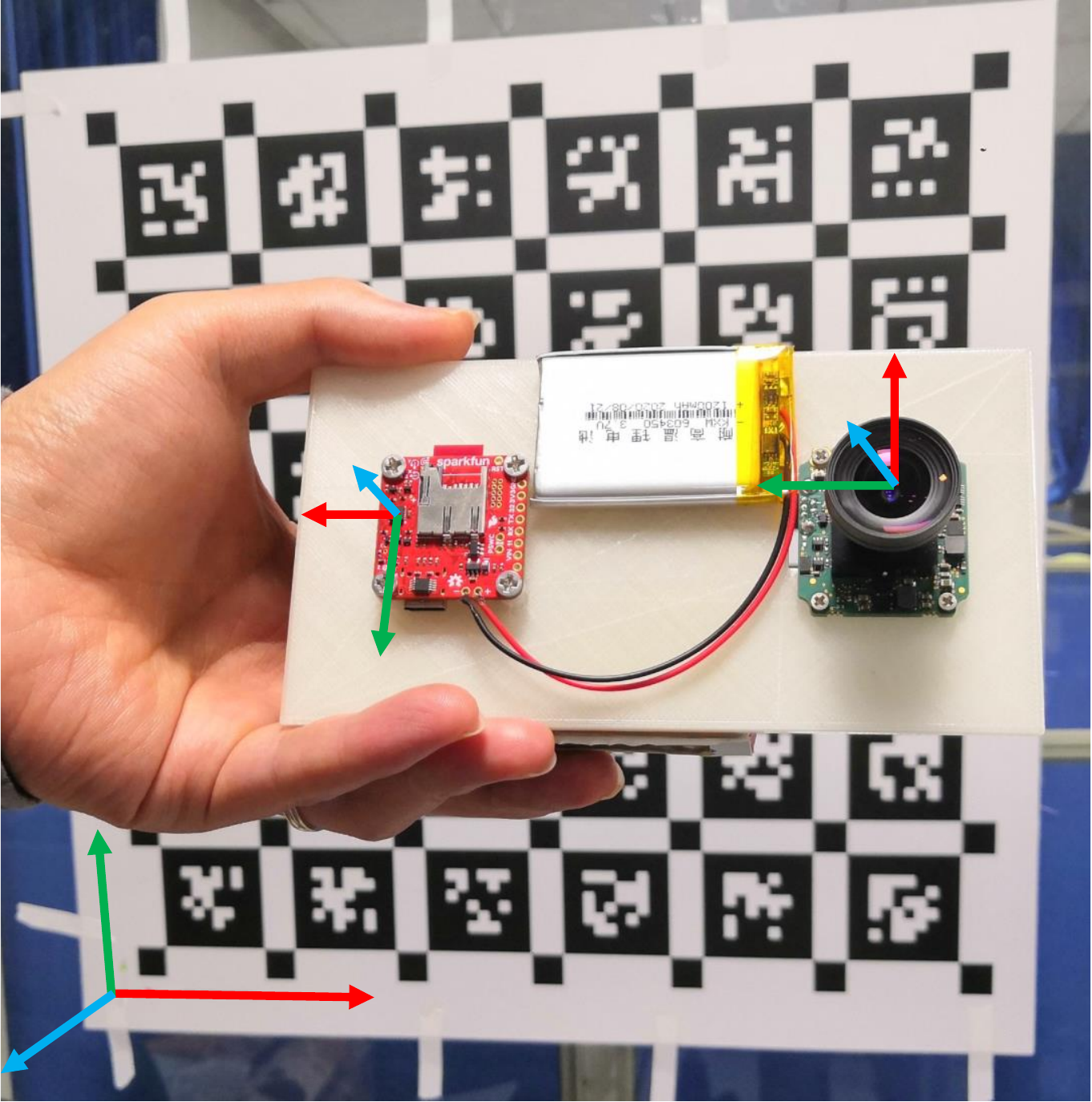}
\caption{Sensor calibration setups: (top) an IDS uEye 3241LE-M-GL camera and a SparkFun OpenLog Artemis IMU board, 
(bottom) a Bluefox MLC202dG camera and the Artemis board.
Also inset in each picture are the coordinate frames for the camera $\{C\}$, 
the IMU $\{I\}$, and the calibration target $\{W\}$.
}
\label{fig:ueye-artemis-20210314}
\end{figure}

\section{Related work}
The wide application range of RS cameras has drawn much research about its effect on a variety of vision-based tasks.
In general, modeling the RS effect leads to better performance with RS cameras.
This improvement has been validated for video stabilization \cite{ringabyEfficient2012},
structure from motion \cite{hedborgRolling2012, saurerSparse2016, ovrenTrajectory2019, daiRolling2016},
dense mapping \cite{kerlDense2015},
vision-aided odometry \cite{liVisionaidedInertialNavigation2014, patron-perezSplinebased2015, schubertRollingshutter2019}.

Though many sensor systems use RS cameras, \eg, Kinect Azure, camera-IMU systems on smartphones, 
few research has been conducted on the spatiotemporal calibration of a RS camera.
Many existing extrinsic calibration studies either deal with GS cameras or assume that the data are 
captured when the sensor system is held still.
Spatiotemporal calibration for a GS camera-IMU system or a GS camera-lidar system is 
achieved by continuous-time B-splines in  \cite{rehderGeneralApproachSpatiotemporal2016}.
Calibration methods for a GS camera-lidar system with motion relative to the scene have been 
presented in \cite{parkSpatiotemporal2020, nowickiSpatiotemporal2020}.
A camera-lidar system was calibrated with data collected by holding the device at different poses in \cite{kangAutomatic2020}.
Basso \etal \cite{bassoRobust2018} estimated the extrinsic parameters of a RGB-D system using synchronized data at standstills.
To calibrate the line delay of a RS camera, a continuous-time optimization method based on B-splines was
proposed in \cite{othRolling2013}.

Driven by the question of how the RS affects the spatiotemporal calibration, 
we develop a calibration approach for the camera-IMU system and 
quantify the RS effect on sensor assemblies with precise reference values.
Cumulative cubic B-splines were used in \cite{patron-perezSplinebased2015} for
visual inertial SLAM with a RS camera.
They showed that the SLAM system could improve the scale estimation by considering the RS effect.
A study close to ours \cite{leeCalibration2018} presents a discrete-time optimization approach
to calibrate the spatiotemporal parameters of a RS camera-IMU system.
The approach assumes constant IMU biases and linearly interpolates poses for rolling shutter observations from poses at
discrete times. Due to varying time offset in iterations of the nonlinear refinement, the IMU factors have to be integrated repeatedly,
adding to the computation.
A bit surprisingly, their tests showed that their approach and the GS camera-IMU calibration tool in Kalibr
achieved similar spatiotemporal calibration results.

\section{Continuous-time rolling-shutter-camera-IMU calibration}
\label{sec:method}
This section presents our spatiotemporal calibration method for the combined device of a RS camera and an IMU.
The inputs are images of a calibration target, \eg, an Aprilgrid, and the corresponding gyroscope and accelerometer data.
Observations extracted from this data are used in a least-squares solver for estimating the 
spatiotemporal parameters and the device motion.
Since landmark observations in an image are exposed at different times due to the RS effect,
we have to interpolate the camera pose at these observations by fitting motion with sparse poses.
Early studies considering the RS effect often linearly interpolated the camera pose with the constant velocity assumption.
An alternative approach is to use continuous-time basis splines, \eg, \cite{furgaleContinuoustime2012, patron-perezSplinebased2015},
which has better fitting capacity and flexibility with high-order curves.
Our method uses the continuous-time vector-valued B-splines \cite{furgaleContinuoustime2012} to accommodate landmark observations at different times.
In the following, we first briefly review the vector-valued B-splines for representing poses,
and then describe the observation models used in calibration.

\subsection{Continuous-time B-splines}
With B-splines of order $k$, a state variable $\mathbf{v}\in\mathcal{R}^D$ is interpolated by
weighting $N$ control points $\mathbf{v}_i \in \mathcal{R}^D$, 
$i = \begin{bmatrix}
0 & 1 & \cdots & N-1
\end{bmatrix}$,
\begin{equation}
\mathbf{v}(t) = \sum_{j=0}^{N-1} \mathbf{v}_j B_{j,k}(t).
\end{equation}
The weights $B_{j,k}(t)$ of these control points are computed with basis splines 
which are analytical functions of time defined recursively \cite{qinGeneral26},
\begin{subequations}\label{eq:Bspline}
	\begin{align}
	B_{i,1}(t)&=\begin{cases}
	1, t\in[t_i, t_{i+1})\\
	0, t\notin[t_i, t_{i+1})
	\end{cases}\\
	B_{j,k}(t)&=\frac{t-t_j}{t_{j+k-1}-t_j}B_{j,k-1}(t)+\frac{t_{j+k}-t}{t_{j+k}-t_{j+1}}B_{j+1,k-1}(t).
	\end{align}
\end{subequations}
where the epochs denoted by $\{t_i\}$ are also known as knots.
A spline takes nonzero values in only $k$ time intervals.
Thus, the vector variable $\mathbf{v}(t)$ fitted by B-splines of order $k$ is determined by $k$ 
control points, $\mathbf{v}_j, \mathbf{v}_{j+1}, \dots, \mathbf{v}_{j+k-1} \in \mathcal{R}^D$,
that contribute to its value at time $t \in [t_j, t_{j+1})$:
\begin{equation}
\mathbf{v}(t) = \sum_{i=j}^{j+k-1}\mathbf{v}_i B_{i,k}(t).
\end{equation}
Defining $u = (t - t_j) / (t_{j+1} - t_j)$, 
$\mathbf{v}(t)$ can be written in matrix form \cite{qinGeneral26},
which facilitates efficient value and derivative evaluation,
\begin{equation}
\mathbf{v}(t) = \begin{bmatrix}
\mathbf{v}_j &  \mathbf{v}_{j+1} & \cdots & \mathbf{v}_{j+k-1}
\end{bmatrix} M^{(k)} \mathbf{u}
\end{equation}
where $\mathbf{u} = \begin{bmatrix}1 & u & \cdots & u^{k-1} \end{bmatrix}^\intercal$.
The entries of $k\times k$ matrix $M^{(k)}$ can be found in \cite{qinGeneral26}.
For uniform B-splines of evenly spaced knots, its entries $m_{i,j}$ can be computed analytically by
\begin{equation}
m_{i, j} = \frac{C_{k-1}^j}{(k-1)!} \sum_{s=i}^{k-1}(-1)^{s-i} C_k^{s-i} \cdot 
(k-1-s)^{k-1-j}
\end{equation}
with $i, j \in \begin{bmatrix}
0 & 1 & \cdots & k-1
\end{bmatrix}$ and the binomial coefficient 
$C_n^i = n! / (i!(n-i)!)$.

To fit motion with vector-valued B-splines, 
the device pose $\mathbf{T}_{WI} = \begin{bmatrix}
\mathbf{R}_{WI} & \mathbf{p}_{WI}
\end{bmatrix}$ relative to the $\{W\}$ frame on the calibration target
is expressed by the angle-axis representation $\boldsymbol{\phi} \in \mathcal{R}^3$ for $\mathbf{R}_{WI}$ 
and the translation component $\mathbf{p}_{WI}(t)$.
Thus, the motion spline with $N$ control points is given by 
\begin{equation}
	\begin{bmatrix}
	\boldsymbol{\phi}(t) \\
	\mathbf{p}(t)
	\end{bmatrix} = \sum_{j=0}^{N-1}
	\begin{bmatrix}
	\boldsymbol{\phi}_j \\
	\mathbf{p}_j
	\end{bmatrix} B_{j,k}(t)
	\label{eq:pose_spline}
\end{equation}
where $\boldsymbol{\phi}_j$ and $\mathbf{p}_j$ are the control points for rotation and translation, respectively.

The above describes the vector-valued B-splines for which the control points are in a vector space.
It is also possible to define cumulative B-splines on Lie groups, \eg, SO(3) \cite{sommerEfficient2020}.
The cumulative B-splines on SO(3) are free of rotation singularity, but are more complex when dealing with unevenly spaced knots.
We choose to use the vector-valued B-splines, partly
for fair comparison to existing methods.
To deal with potential angle jumps (axis flips) at 2$k\pi$, $k=[1, 2, \cdots]$, in the angle-axis representation,
the approach described in \cite{othRolling2013} is used to ensure consecutive angle-axis values are close.

To model the time-varying IMU biases $\mathbf{b}(t)$, we fit them by B-splines with $M$ control points $\mathbf{b}_j$, $j = [0, 1, \cdots, M-1]$,
\begin{equation}
\mathbf{b}(t) = [\mathbf{b}_g^\intercal(t) \enskip \mathbf{b}_a^\intercal(t)]^\intercal = 
\sum_{j=0}^{M-1} \mathbf{b}_{j} B_{j,k}(t)
\end{equation}
where $\mathbf{b}_g$ and $\mathbf{b}_a$ are gyroscope and accelerometer biases, respectively.

\subsection{Observations}
\label{subsec:observations}
For the camera-IMU system, the camera provides images from which landmark observations are detected
and the IMU provides angular velocity and linear acceleration measurements.

The landmark observations are modeled with the classic reprojection model.
For a landmark $k$ of homogeneous coordinates $\mathbf{l}_k^W$,
its reprojection error in image $j$ of timestamp $t_j$ per the camera clock is
\begin{equation}
\label{eq:reprojection}
\mathbf{r}_{jk} = \mathbf{h}(\mathbf{T}_{CI}[\mathbf{T}_{WI}(t_{j} + t_{IC} + vd)]^{-1} \mathbf{l}_k^W) - \mathbf{z}_{jk} + \mathbf{n}_c,
\end{equation}
where $\mathbf{h}(\cdot)$ is the camera projection model with the intrinsic parameters $\boldsymbol{\theta}$, 
the camera extrinsic parameters are in $\mathbf{T}_{CI}$,
the observation is $\mathbf{z}_{jk} = [u, v]^\intercal$, 
and the observation time $t_j + t_{IC} + vd$ accounts for 
the camera time offset $t_{IC}$ relative to the IMU and line delay $d$.
The noise $\mathbf{n}_c$ affecting image observations is assumed
to be 2D white Gaussian with magnitude $\sigma_c$ in each dimension.

We adopt two IMU models presented in \cite{rehderExtending2016},
the calibrated model with bias terms, and the scale-misalignment model 
considering scale and misalignment effects besides biases.

Recall that the accelerometer triad measures the acceleration in the IMU frame,
$\mathbf{a}_s^I$, due to specific forces,
\begin{equation}
\mathbf{a}_s^I = \mathbf{R}_{WI}^\intercal (\ddot{\mathbf{p}}_{WI} - \mathbf{g}^W).
\end{equation}
The gyroscope measures angular velocity of the device in $\{I\}$ frame, $\boldsymbol{\omega}_{WI}^I$.
Both $\mathbf{a}_s^I$ and $\boldsymbol{\omega}_{WI}^I$ can be expressed with the pose B-spline \eqref{eq:pose_spline} \cite{rehderExtending2016}.

In the calibrated model, the IMU measurements $\mathbf{a}_m$ and
$\boldsymbol{\omega}_m$ are affected by accelerometer and gyroscope biases,
$\mathbf{b}_a$ and $\mathbf{b}_g$, and Gaussian white noise processes,
$\boldsymbol{\nu}_{a}$ and $\boldsymbol{\nu}_{g}$,
\begin{equation}
\begin{split}
\mathbf{a}_m &= \mathbf{a}_s^I + \mathbf{b}_a + \boldsymbol{\nu}_a \\
\boldsymbol{\omega}_m &= \boldsymbol{\omega}_{WI}^I +
\mathbf{b}_g + \boldsymbol{\nu}_g
\end{split}
\label{eq:calibrated_model}
\end{equation}
The biases are usually assumed to be driven by Gaussian white
noise processes, $\boldsymbol{\nu}_{ba}$ and $\boldsymbol{\nu}_{bg}$,
\begin{equation}
\dot{\mathbf{b}}_a = \boldsymbol{\nu}_{ba} \quad \dot{\mathbf{b}}_g = \boldsymbol{\nu}_{bg}.
\end{equation}
The power spectral densities of $\boldsymbol{\nu}_{a}$, $\boldsymbol{\nu}_{g}$, 
$\boldsymbol{\nu}_{ba}$, and $\boldsymbol{\nu}_{bg}$, 
are usually assumed to be $\sigma^2_a\mathbf{I}_3$, $\sigma^2_g \mathbf{I}_3$, $\sigma^2_{ba}\mathbf{I}_3$, and $\sigma^2_{bg}\mathbf{I}_3$, respectively.

For the scale-misalignment model, the accelerometer measurement $\mathbf{a}_m$ is corrupted by
systematic errors encoded in a lower triangular $3\times 3$ matrix $\mathbf{M}_a$, $\mathbf{b}_a$, and $\boldsymbol{\nu}_a$,
\begin{equation}
\mathbf{a}_m = \mathbf{M}_a \mathbf{a}_s^I + \mathbf{b}_a + \boldsymbol{\nu}_a.
\label{eq:accel_scale_misalign}
\end{equation}
The 6 nonzero entries of $\mathbf{M}_a$ encompass 3-DOF scale factor error and 3-DOF misalignment.
The gyroscope measurement $\boldsymbol{\omega}_m$ is corrupted by systematic errors encoded in a 
$3\times 3$ matrix $\mathbf{M}_g$ and the $g$-sensitivity effect encoded in a $3\times 3$ matrix $\mathbf{M}_s$,
$\mathbf{b}_g$, and $\boldsymbol{\nu}_g$,
\begin{equation}
\boldsymbol{\omega}_m = \mathbf{M}_g \boldsymbol{\omega}_{WI}^I +
\mathbf{M}_s \mathbf{a}_s^I + 
\mathbf{b}_g + \boldsymbol{\nu}_g
\label{eq:gyro_scale_miaslign}
\end{equation}
The 9 entries of $\mathbf{M}_g$ encompass 3-DOF scale factor error, 3-DOF misalignment,
and 3-DOF relative orientation between the gyroscope input axes and the $\{I\}$ frame defined by the accelerometer input axes.

In summary, variables to be estimated and known parameters in our optimization-based calibration are listed in Table~\ref{tab:variables}.

\begin{table}[]
\caption{The variables and known parameters in the RS camera-IMU calibration problem.}
\label{tab:variables}
	\begin{tabular}{ll}
		\hline\hline
		\multicolumn{2}{c}{Estimated variables}                                                                                                   \\ \hline
		& Time-varying                                                                                                     \\ \hline
		$\mathbf{T}_{WI}$     & \begin{tabular}[c]{@{}l@{}}pose of the IMU relative to the target\\ expressed by sixth-order B-splines\end{tabular} \\ \hline
		$\mathbf{b}$          & \begin{tabular}[c]{@{}l@{}}gyroscope and accelerometer biases\\ expressed by sixth-order B-splines\end{tabular}        \\ \hline
		& Time-invariant                                                                                                    \\ \hline
		$\mathbf{T}_{CI}$     & pose of the camera relative to the IMU                                                                            \\ \hline
		$t_{IC}$              & camera time offset relative to the IMU                                                                            \\ \hline
		$\mathbf{g}^W/g$        & gravity direction in the target frame                                                                             \\ \hline
		$d$                   & line delay of the rolling shutter camera                                                                          \\ \hline \hline
		\multicolumn{2}{c}{Known parameters}                                                                                                      \\ \hline
		$\mathbf{l}_k^W$      & landmark coordinates on the target                                                                                           \\ \hline
		$\boldsymbol{\theta}$ & camera intrinsic parameters including distortion                                                                  \\ \hline
		$g$                   & local gravity magnitude, \eg, 9.80665                                                                      \\ \hline \hline
	\end{tabular}
\end{table}

\section{Experiments}
\label{sec:experiments}
To study the RS effect on spatiotemporal calibration of a camera-IMU system and 
validate the proposed calibration method,
we conducted simulation with four public calibration datasets and 
tests on real data captured by two industrial camera-IMU units.

The proposed calibration approach was implemented on top of Kalibr \cite{rehderExtending2016}.
For the following tests, both poses and IMU biases were fitted by sixth-order B-splines which are
piece-wise fifth-degree polynomials,
accommodating the required diverse motion to render the extrinsic parameters observable.
Knots were placed at 100 Hz for pose B-splines, and at 50 Hz for bias B-splines.
Both simulation and calibration adhered to image noise $\sigma_c$ = 1 pixel and pinhole camera model with equidistant distortion.
The equidistant distortion model encoded by four parameters was chosen for 
its fitness to a wide range of lenses \cite{kannalaGeneric2006} and its high precision \cite{usenkoDouble2018}.
The camera intrinsic parameters required by the spatiotemporal calibration were
obtained with the camera calibration tool in Kalibr \cite{mayeSelfsupervised2013} when necessary.

To evaluate calibration results, differences between reference values and estimated parameters are computed.
For the extrinsic parameter $\mathbf{T}_{CI} = \begin{bmatrix}
\mathbf{R} & \mathbf{p}
\end{bmatrix}$, 
deviation of its estimated value $\bar{\mathbf{T}}_{CI} = \begin{bmatrix}
\bar{\mathbf{R}} & \bar{\mathbf{p}}
\end{bmatrix}$ is computed by
\begin{equation}
\Delta \mathbf{R} = \mathbf{R}^\intercal \bar{\mathbf{R}} \quad
\Delta \mathbf p = \mathbf{R}^\intercal (\bar{\mathbf{p}} - \mathbf p).
\end{equation}
And its estimation error is quantified by the angle of the angle-axis representation of $\Delta \mathbf{R}$, 
$\norm{\angle(\Delta \mathbf{R})}$, and $\norm{\Delta \mathbf{p}}$.

\subsection{Simulation}
To quantify how RS affects the spatiotemporal parameters of a camera-IMU system,
we simulated RS camera and IMU data based on four public calibration datasets, 
and then estimated these parameters with the proposed method.

The four public datasets include the camera-IMU calibration sample provided by Kalibr,
and the calibration data of the EuRoC, TUM-VI, and UZH datasets.
The four calibration datasets are summarized here
\footnote{\url{https://github.com/VladyslavUsenko/basalt-mirror/blob/master/doc/Calibration.md}}.

To create simulated data, for every dataset,
we first ran the GS-camera-IMU calibration tool in Kalibr and saved the fitted pose B-splines.
Then from the B-splines, noisy RS camera observations of the target and IMU measurements spanning 50 seconds were
simulated using reference camera and IMU parameters for the dataset.
These reference parameters were obtained by simply rounding the calibrated precise values for easy interpretation.
For all datasets, the IMU noise parameters in simulating IMU data and in the subsequent calibration were 
identical to those for ADIS16448 found in the Kalibr sample dataset, \ie, 
accelerometer noise density $\sigma_a$ = 1.0$\cdot10^{-2}$ $m/s^2/\sqrt{Hz}$, 
accelerometer random walk $\sigma_{ba}$ = 2.0$\cdot10^{-4}$ $m/s^3/\sqrt{Hz}$, 
gyroscope noise density $\sigma_g=5.0\cdot10^{-3}$ $rad/s/\sqrt{Hz}$, 
gyroscope random walk $\sigma_{bg}=4.0\cdot10^{-6}$ $rad/s^2/\sqrt{Hz}$.

The RS image observations were simulated at four line delays, 137.5, 82.5, 51.563, and 41.25 ms,
which correspond to pixel clocks, 12, 20, 32, and 40 MHz, of a Matrix Vision Bluefox MLC202dG camera with a line length 1650 pixels.
To avert a bias in the camera time offset when a GS-camera-IMU calibration method processes the simulated RS camera data, 
a simulated RS image was assigned timestamp of its central row.

\begin{figure}[t]
\centering
\includegraphics[width=0.8\linewidth]{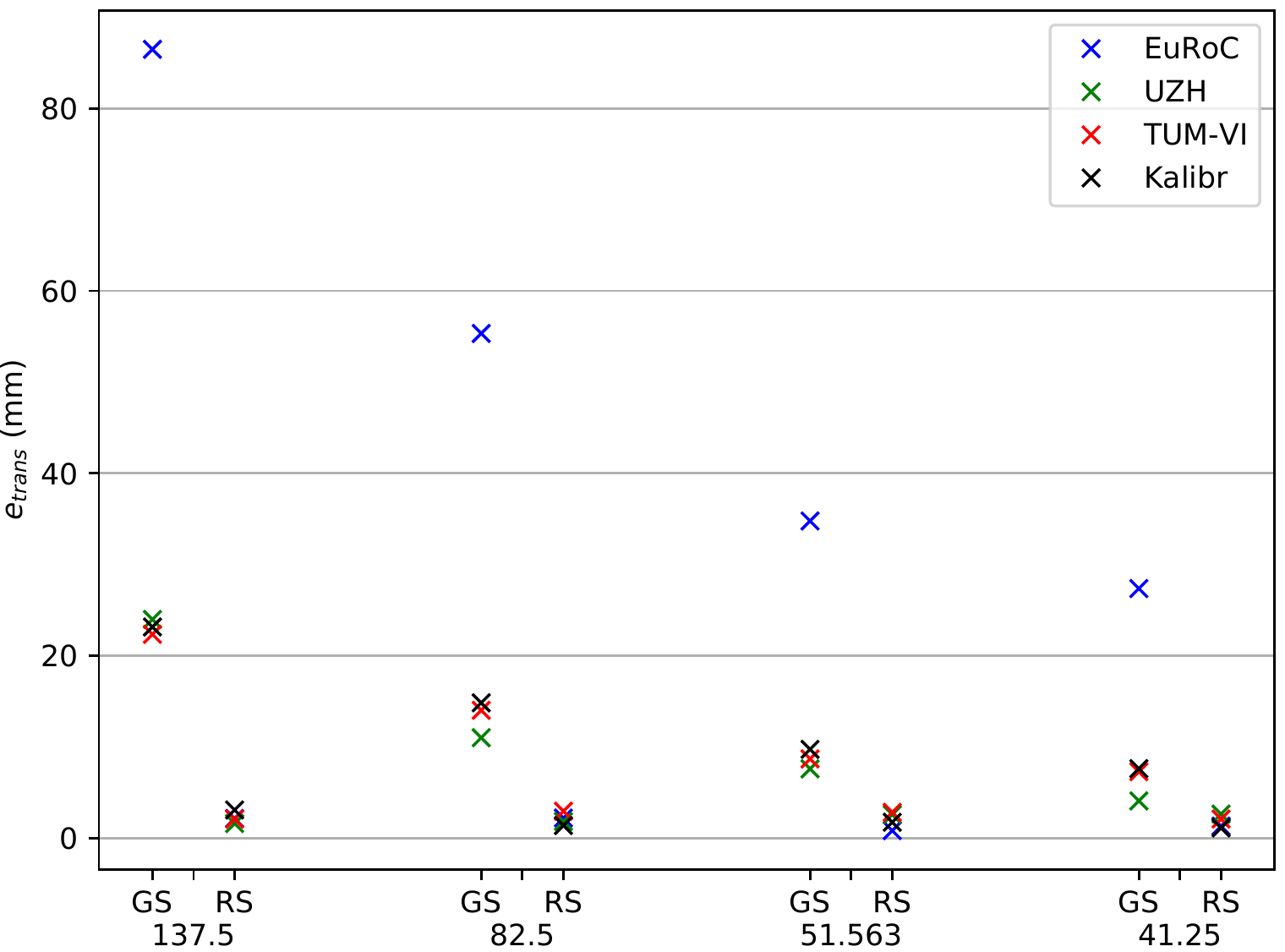} \\
\includegraphics[width=0.82\linewidth]{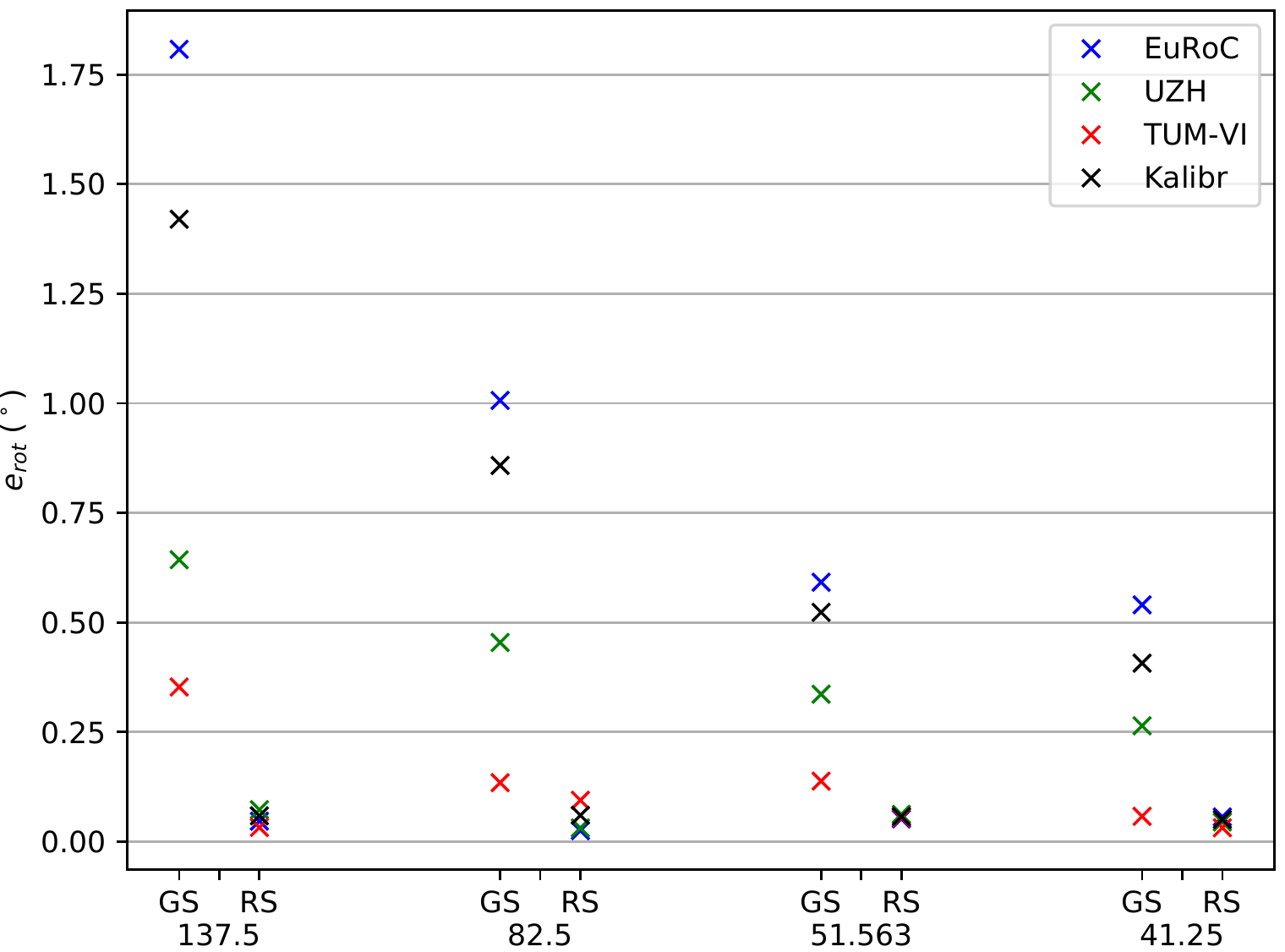}
\caption{Translation (top) and rotation (bottom) errors of the Kalibr camera-IMU calibration method (denoted by GS) 
and the proposed method for RS cameras (denoted by RS) with simulated data from four public calibration datasets, EuRoC, UZH, TUM-VI, and Kalibr.
The simulation used four line delays, 137.5, 82.5, 51.563, and 41.25 $\mu s$.
}
\label{fig:sim_error_pose}
\end{figure}

\begin{figure}[t]
\centering
\includegraphics[width=0.82\linewidth]{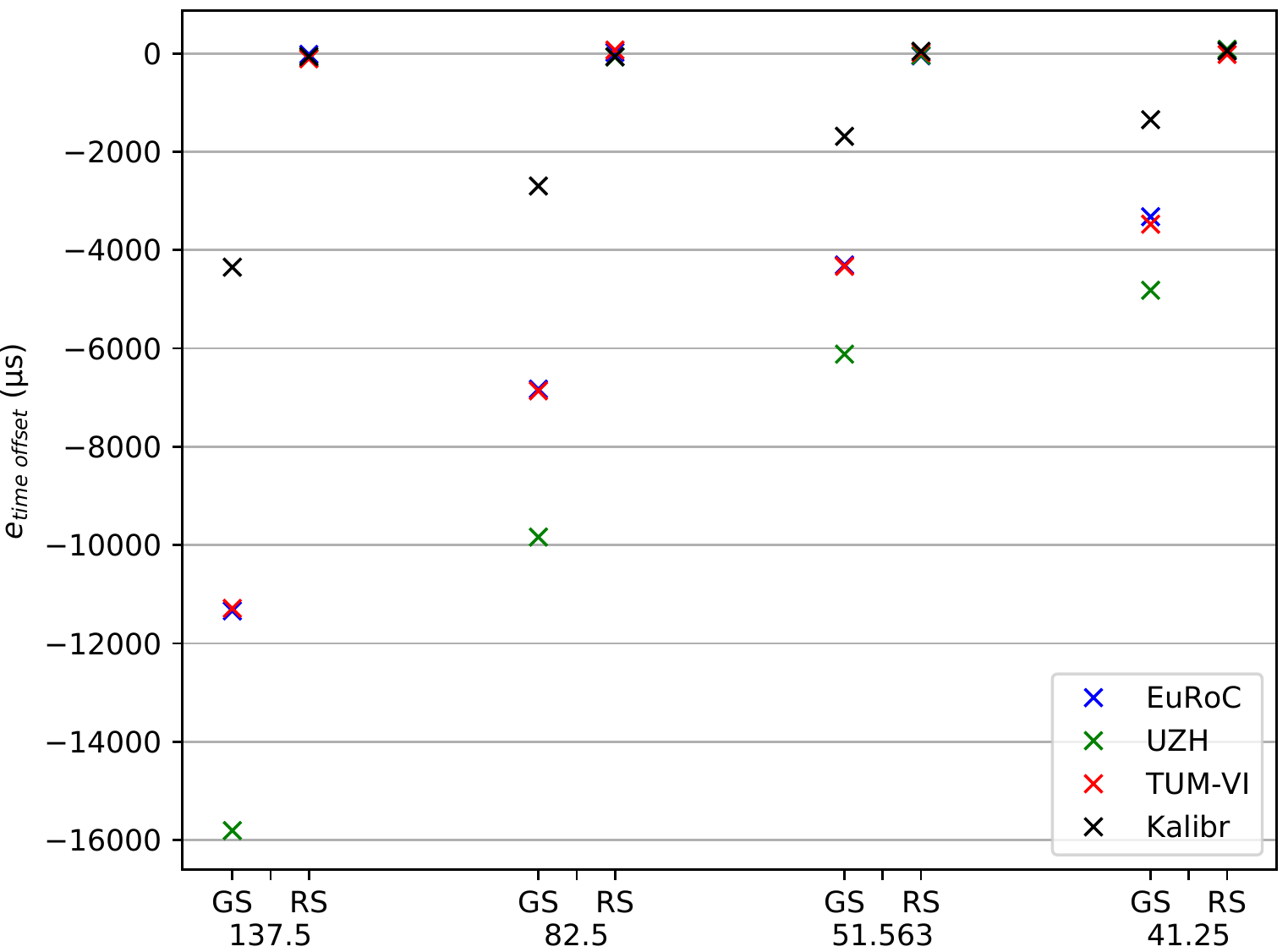} \\ (a) \\
\includegraphics[width=0.8\linewidth]{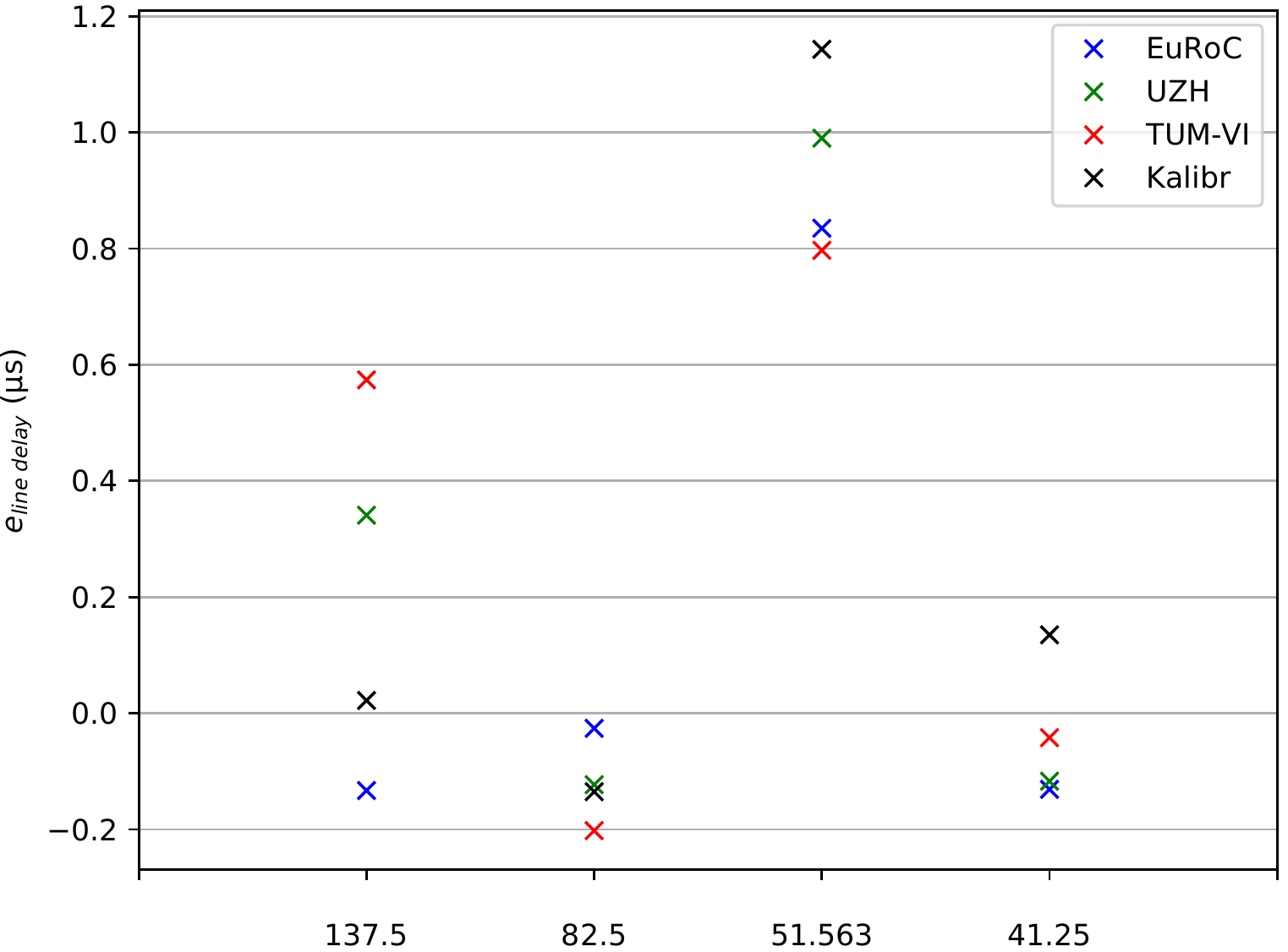} \\ (b)
\caption{Top: Camera time offset errors of the Kalibr camera-IMU calibration method (denoted by GS) and 
the proposed method for RS cameras (denoted by RS) on the simulated data. 
Bottom: Line delay errors of the proposed method on the simulated data.
The data were simulated from four public calibration datasets, EuRoC, UZH, TUM-VI, and Kalibr, 
with four line delays, 137.5, 82.5, 51.563, and 41.25 $\mu s$.
}
\label{fig:sim_error_time}
\end{figure}

In the end, the simulated data were processed by our RS camera-IMU calibration method and 
the camera-IMU calibration tool in Kalibr \cite{rehderExtending2016}. 
For both methods, the calibrated IMU model \eqref{eq:calibrated_model} is used.
The errors in $\mathbf{T}_{CI}$, $t_{IC}$ and $d$,
are visualized in Figs. \ref{fig:sim_error_pose}, \ref{fig:sim_error_time}, respectively.
From Fig.~\ref{fig:sim_error_pose}, we see that longer line delays led to greater extrinsic errors for the GS-camera-IMU calibration method, while 
our method maintained small extrinsic errors across varying line delays and datasets.
The GS-camera-based method often had errors more than 0.5$^\circ$ in orientation and 2 cm in translation.
Fig.~\ref{fig:sim_error_time} shows that ignoring the RS effect often 
led to time offset errors greater than 4 ms.
And Fig.~\ref{fig:sim_error_time}(b) shows that line delay could usually be accurately estimated within 1 $\mu$s.
Overall, the simulation shows that the RS effect can noticeably deteriorate the spatiotemporal calibration of a camera-IMU system if ignored, 
and our method can accurately estimate the RS line delay and remove the adverse effect.

\subsection{Real data tests}
We also tested the proposed method on two RS camera-IMU systems built with industrial cameras of delicate RS control.
One device combined an IDS uEye 3241LE-M-GL camera fitted with a Lensagon BM4018S118 lens of 126$^\circ$ diagonal FOV and a SparkFun OpenLog Artemis IMU board.
The other was composed of a Matrix Vision Bluefox MLC202dG camera fitted with a E1M3518 lens of 90$^\circ$ diagonal FOV and an OpenLog Artemis IMU board.
The uEye camera allows switching between RS mode and GS mode in operation.
Thus, the extrinsic calibration result with data captured in GS mode can serve as an accurate reference for $\mathbf{T}_{CI}$.
The Artemis board carries an InvenSense ICM-20948 IMU (cost less than \$4), 
and is able to capture the inertial data at about 230 Hz.
The Bluefox camera only supports RS mode, but it has precise pixel clocks and 
a known line length for determining the reference line delays.

In data acquisition, the exposure time of the two cameras was set to 5 ms to reduce motion blur;
the focus distance of both lenses was about 1.5 meters and remained fixed.
For every device, the data collection began with capturing the camera intrinsic calibration data.
For the uEye camera, a video was captured while the camera in GS mode was moved in front of the static target.
For the Bluefox camera, an image sequence was captured by holding the RS camera at 100 different poses.
Afterwards, the RS camera-IMU data were collected while the device was moved in front of the static target.
For every device, a set of five one-minute RS sessions was captured at each of four pixel clocks, 12, 20, 32, and 40 MHz.

The IMU noise parameters were estimated by Allan variance analysis as detailed in the supplementary material.
For the accelerometer and gyroscope noise density,
the Allan analysis obtained values reasonably close to those on the ICM-20948 datasheet.
The noise values read from Allan analysis were obtained for three accelerometers and three gyroscopes, and then 
plugged in the compared calibration methods without inflation.
Specifically, accelerometer noise density $\sigma_a$ = 2.3$\cdot10^{-3}$ $m/s^2/\sqrt{Hz}$, 
accelerometer random walk $\sigma_{ba}$ = 6.5$\cdot10^{-5}$ $m/s^3/\sqrt{Hz}$, 
gyroscope noise density $\sigma_g=2.6\cdot10^{-4}$ $rad/s/\sqrt{Hz}$, 
gyroscope random walk $\sigma_{bg}=4.1\cdot10^{-6}$ $rad/s^2/\sqrt{Hz}$.
It is a bit counter-intuitive that the InvenSense IMU has smaller noise parameters than the more expensive ADIS16448.
One possible reason is that the latter's noise values in the Kalibr sample data had been inflated.
In a preliminary test about whether to inflate the noise parameters, we inflated the noise density and 
random walk parameters by scalars $\alpha$ and $\beta$, respectively, 
ran the proposed calibration method, and iterated the two steps.
The two scalars were grid-searched for the minimum total cost.
In the end, we found that inflating the IMU noise parameters led to smaller reprojection errors but
larger accelerometer and gyroscope errors; 
and extraordinary inflation caused difficulty in convergence of the optimization-based method.
Thus, we chose to use the original noise parameters obtained by the principled method.

For all RS camera and IMU data captured by the two devices, 
we processed them by the proposed method and the GS-camera-IMU calibration tool in Kalibr.
Each method has two variants, one with the calibrated IMU model, 
and the other with the scale-misalignment IMU model (Section~\ref{subsec:observations}).
The two variants aim to tease apart the scale and misalignment effect commonly 
found in a consumer-grade IMU from the extrinsic calibration.
Overall, we have four calibration methods of shorthand names: 
GS calibrated IMU, GS scale-misalign IMU, RS calibrated IMU, RS scale-misalign IMU.

\begin{figure}[]
\centering
\includegraphics[width=0.7\linewidth]{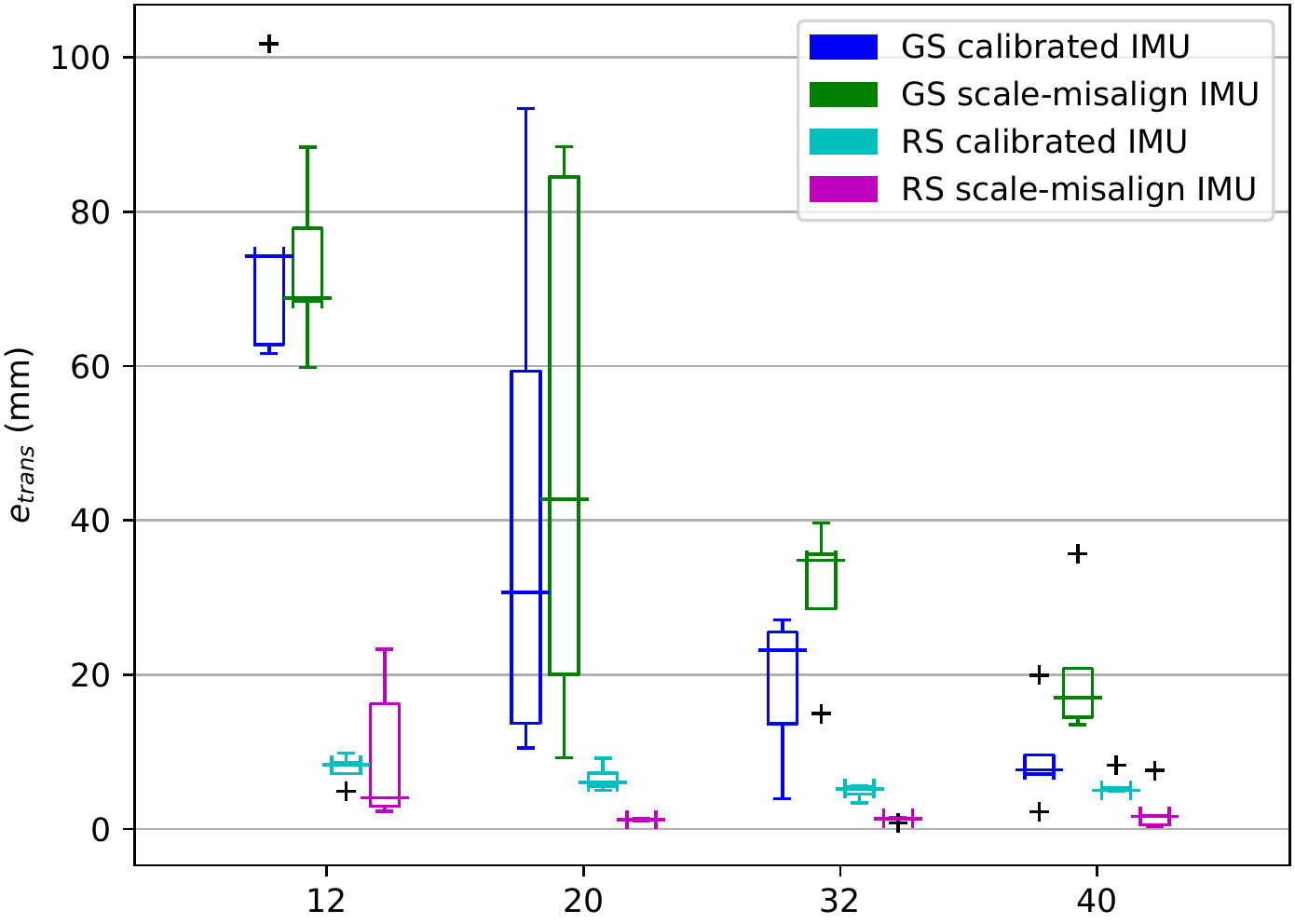} \\(a)\\
\includegraphics[width=0.7\linewidth]{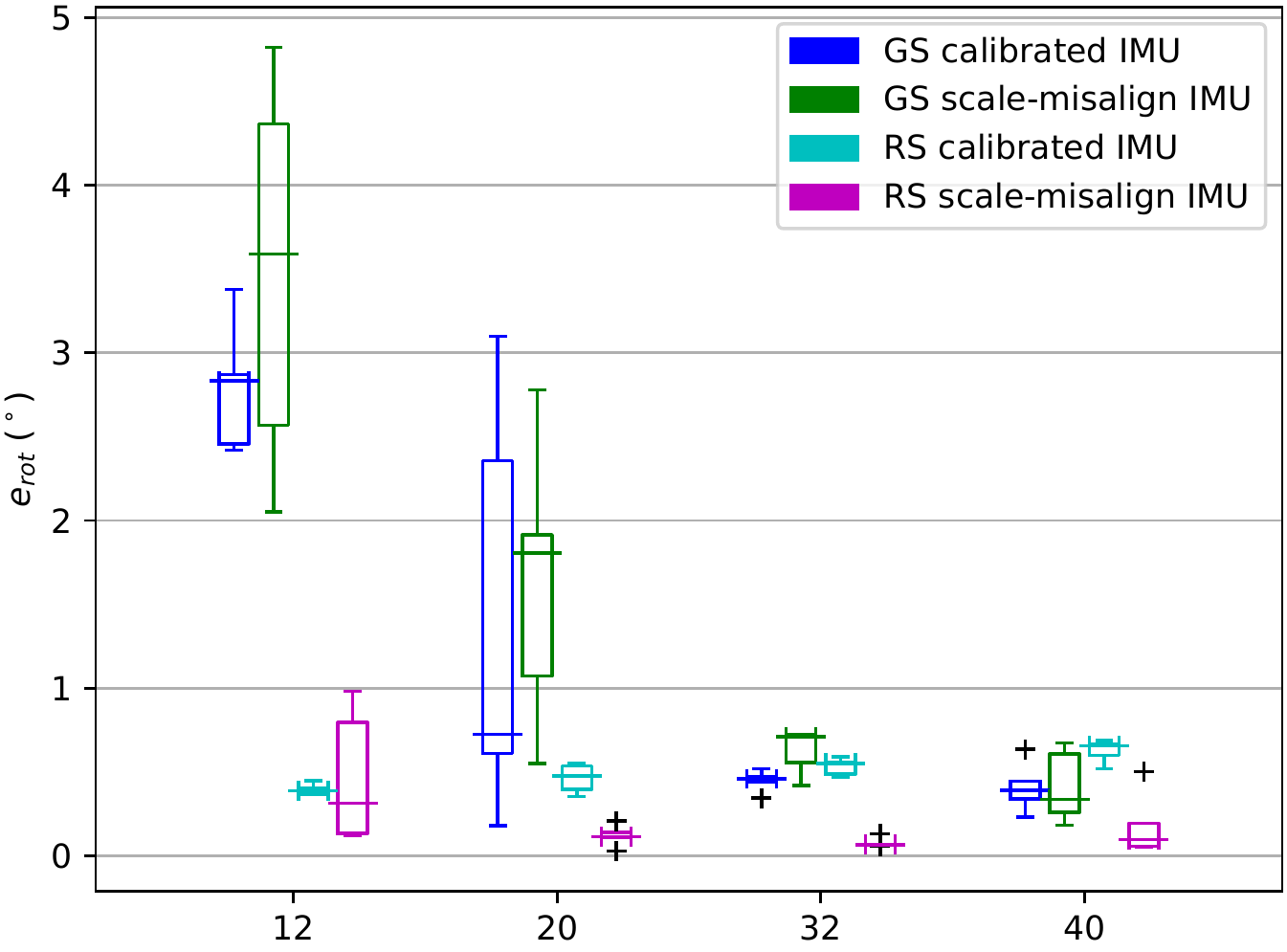} \\(b)\\
\includegraphics[width=0.7\linewidth]{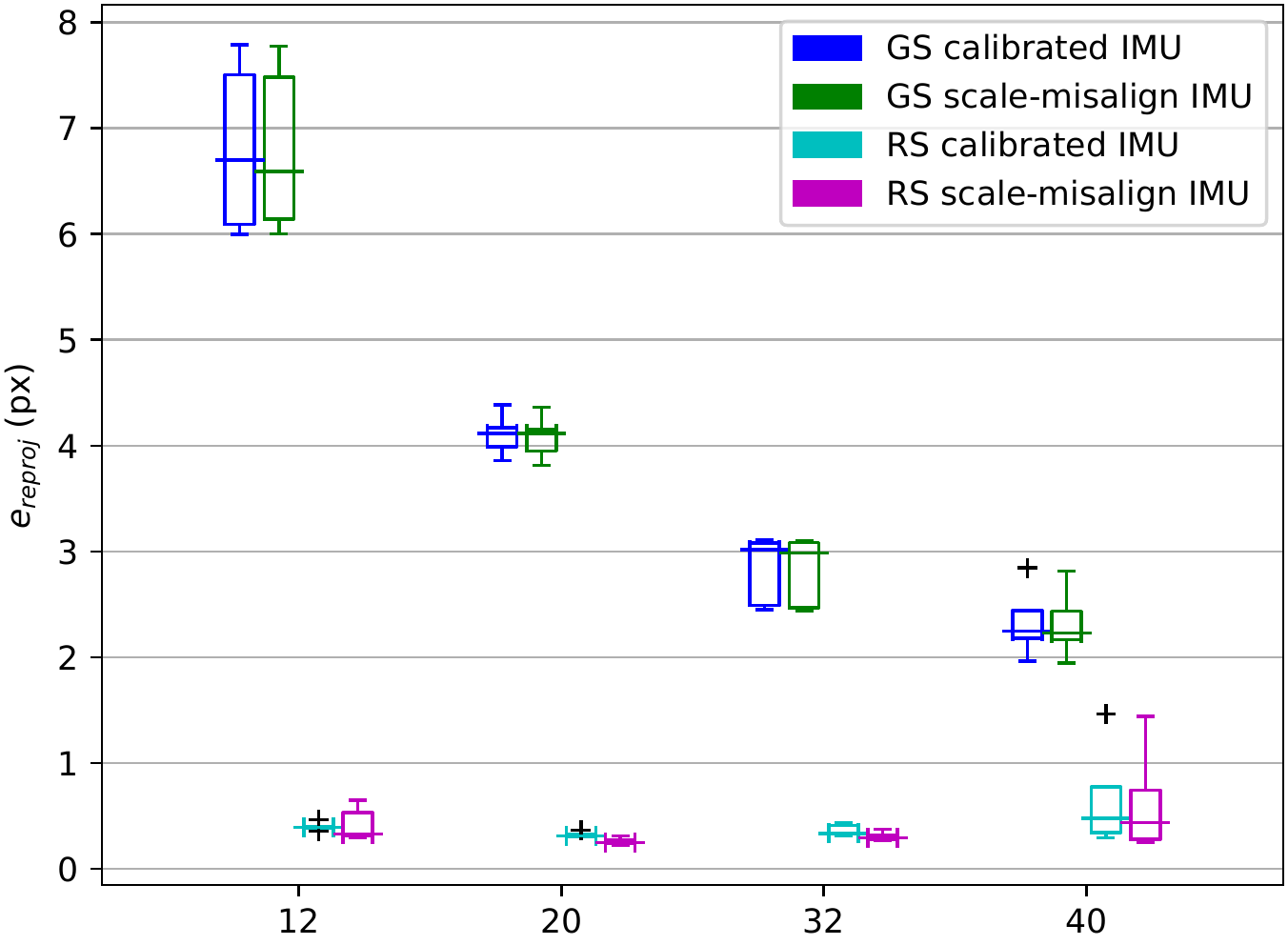} \\ (c)
\caption{
Translation (a) and rotation (b) errors, and median reprojection errors (c) of the proposed method (denoted by RS) and 
the Kalibr camera-IMU calibration tool (denoted by GS) with both calibrated
and scale-misalignment IMU model, on the uEye-Artemis data.
}
\label{fig:ueye-pose-reproj}
\end{figure}

\subsubsection{uEye-Artemis}
\label{subsubsec:ueye}
For the uEye camera and Artemis IMU assembly, to obtain the camera intrinsic parameters, 
the video captured in GS mode were 
down-sampled in frame rate and then processed by the intrinsic calibration tool in Kalibr \cite{mayeSelfsupervised2013}.
The intrinsic parameters were used by all the compared calibration methods.
To obtain the reference extrinsic parameters,
we used a one-minute camera-IMU session captured with the uEye camera in GS mode at the pixel clock 40 MHz.
The session was then processed by the Kalibr camera-IMU calibration tool with the scale-misalignment IMU model, 
attaining the reference extrinsic parameters $\mathbf{T}_{CI}$ 
which agreed very well with values $\tilde{\mathbf{T}}_{CI}$ measured by ruler,
\begin{equation}
\label{eq:ueye_T_CI}
\begin{split}
	\mathbf{T}_{CI} &= \begin{bmatrix}
	-0.999 & -0.032 & 0.014 & -0.015 \\
	-0.032 & 0.999 & 0.024 & 0.061 \\
	-0.015 & 0.023 & -1.000 & -0.020
	\end{bmatrix} \\
    \tilde{\mathbf{T}}_{CI} &= \begin{bmatrix}
    -1 & 0 & 0 & -0.014 \pm 0.003 \\
    0 & 1 & 0 & 0.062 \pm 0.003 \\
    0 & 0 & -1 & -0.024 \pm 0.003
    \end{bmatrix}
\end{split}
\end{equation}
where the uncertainty in a measured distance is about 3 mm.

The RS camera-IMU data were processed by the aforementioned four calibration methods.
The translation, rotation, and reprojection errors are visualized in Fig.~\ref{fig:ueye-pose-reproj}.
Similarly to results in simulation, Fig.~\ref{fig:ueye-pose-reproj}(a) and (b) show that
extrinsic calibration errors grew with line delays for the GS calibration methods, 
and that the RS calibration methods kept relatively small errors across pixel clocks.
For the RS calibration methods, the scale-misalignment IMU model slightly improved the estimated extrinsic parameters.
Overall, the RS calibration methods consistently outstripped the GS-based methods for both IMU models.
For instance, at 40 MHz, the GS calibration methods had errors about 0.4$^\circ$ in rotation and 15 mm in translation,
whereas the RS calibration method with the scale-misalignment IMU
model incurred errors about 0.15$^\circ$ in rotation, and 2 mm in translation.
Fig.~\ref{fig:ueye-pose-reproj}(c) shows that the RS effect caused reprojection errors about 5 pixels for GS calibration methods while
the RS calibration methods had sub-pixel errors.
These results confirm that considering the RS effect could substantially improve the extrinsic estimation.
Considering that the line delays for the uEye camera at pixel clocks, 12, 20, 32, 40 MHz, are roughly 107.7, 64.6, 40.4, and 32.3 $\mu$s, 
the improvements are quite relevant for consumer cameras which often have line delays in range 25 -- 60 $\mu$s \cite{othRolling2013}.

\begin{figure}[]
\centering
\includegraphics[width=0.7\linewidth]{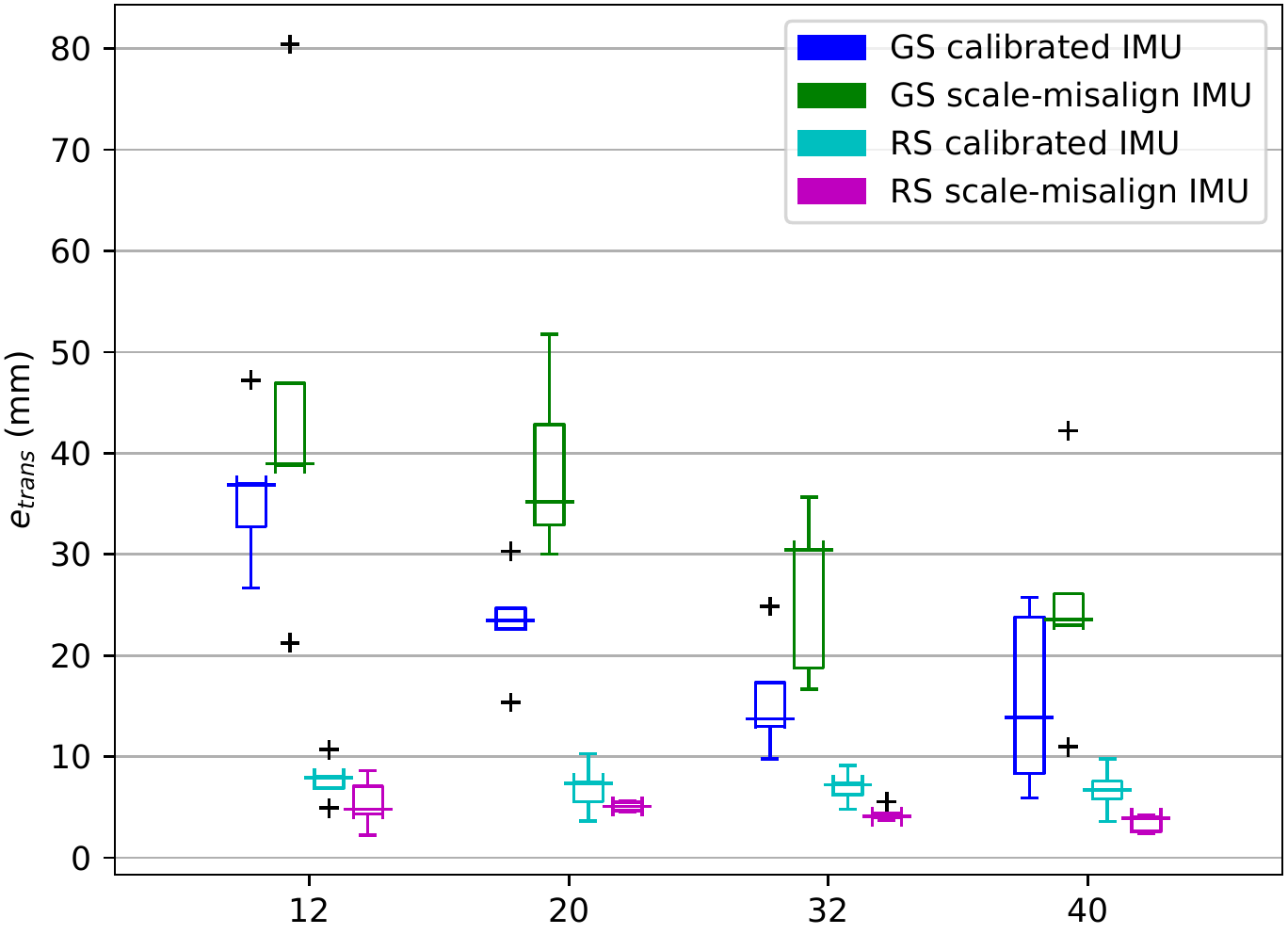} \\(a)\\
\includegraphics[width=0.7\linewidth]{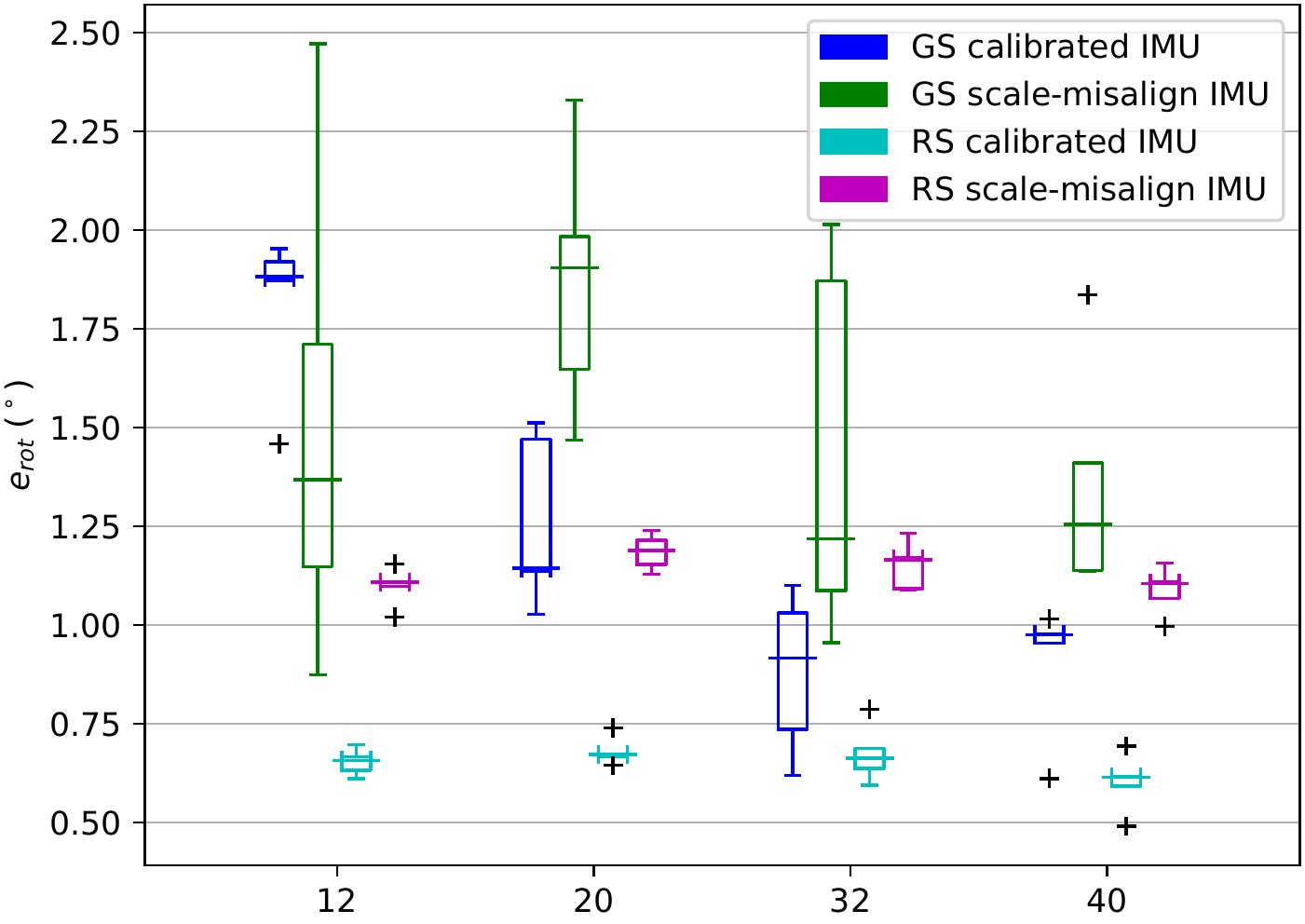} \\(b)\\
\includegraphics[width=0.7\linewidth]{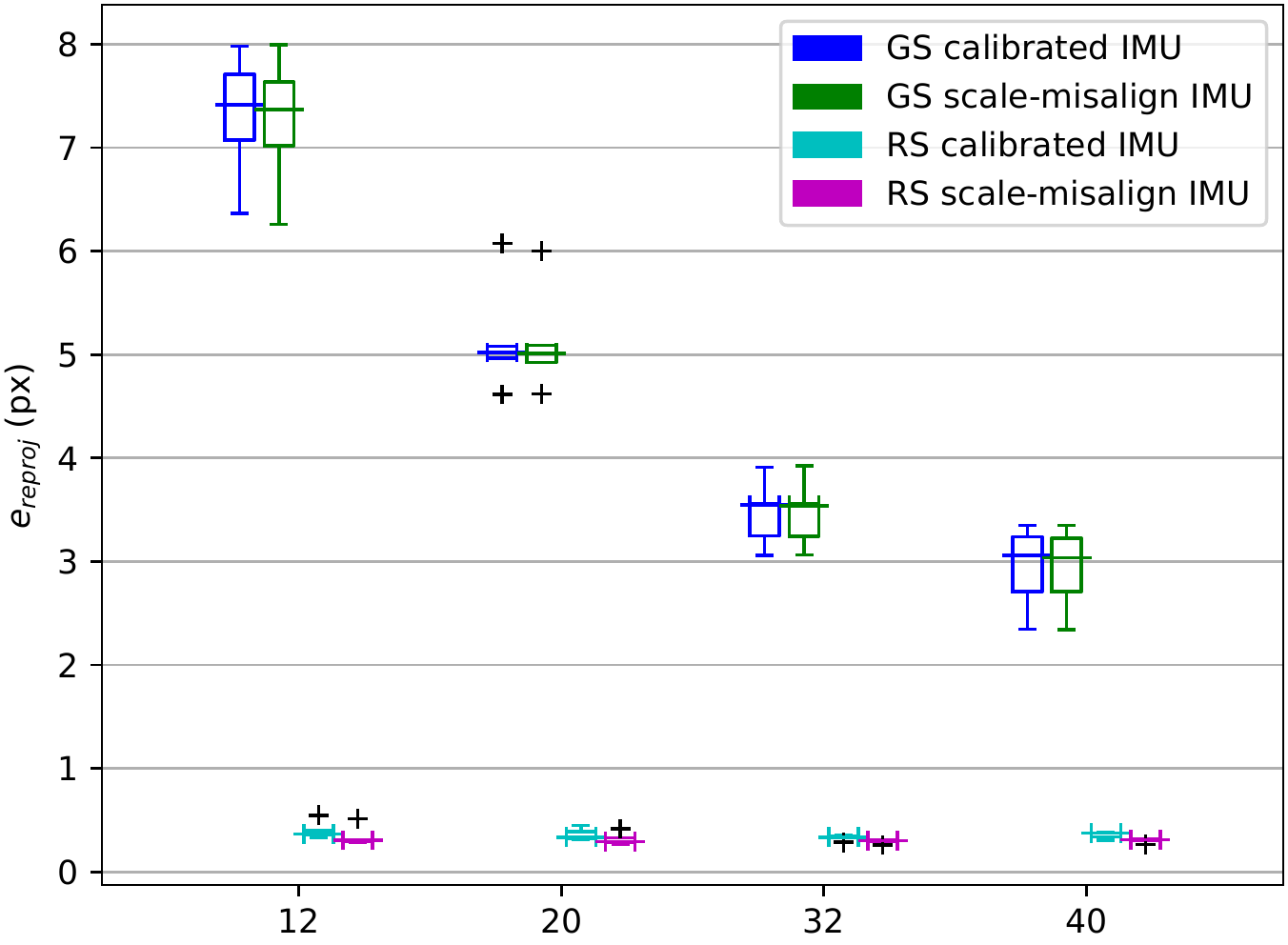} \\ (c)
\caption{
Translation (a), rotation (b) errors, and median reprojection errors (c) of the proposed method (denoted by RS) and 
the Kalibr camera-IMU calibration tool (denoted by GS) with both calibrated
and scale-misalignment IMU model on the Bluefox-Artemis data.
Note that the reference extrinsic parameters were measured by hand.
}
\label{fig:bluefox-pose-reproj}
\end{figure}

\begin{figure}[]
\centering
\includegraphics[width=0.7\linewidth]{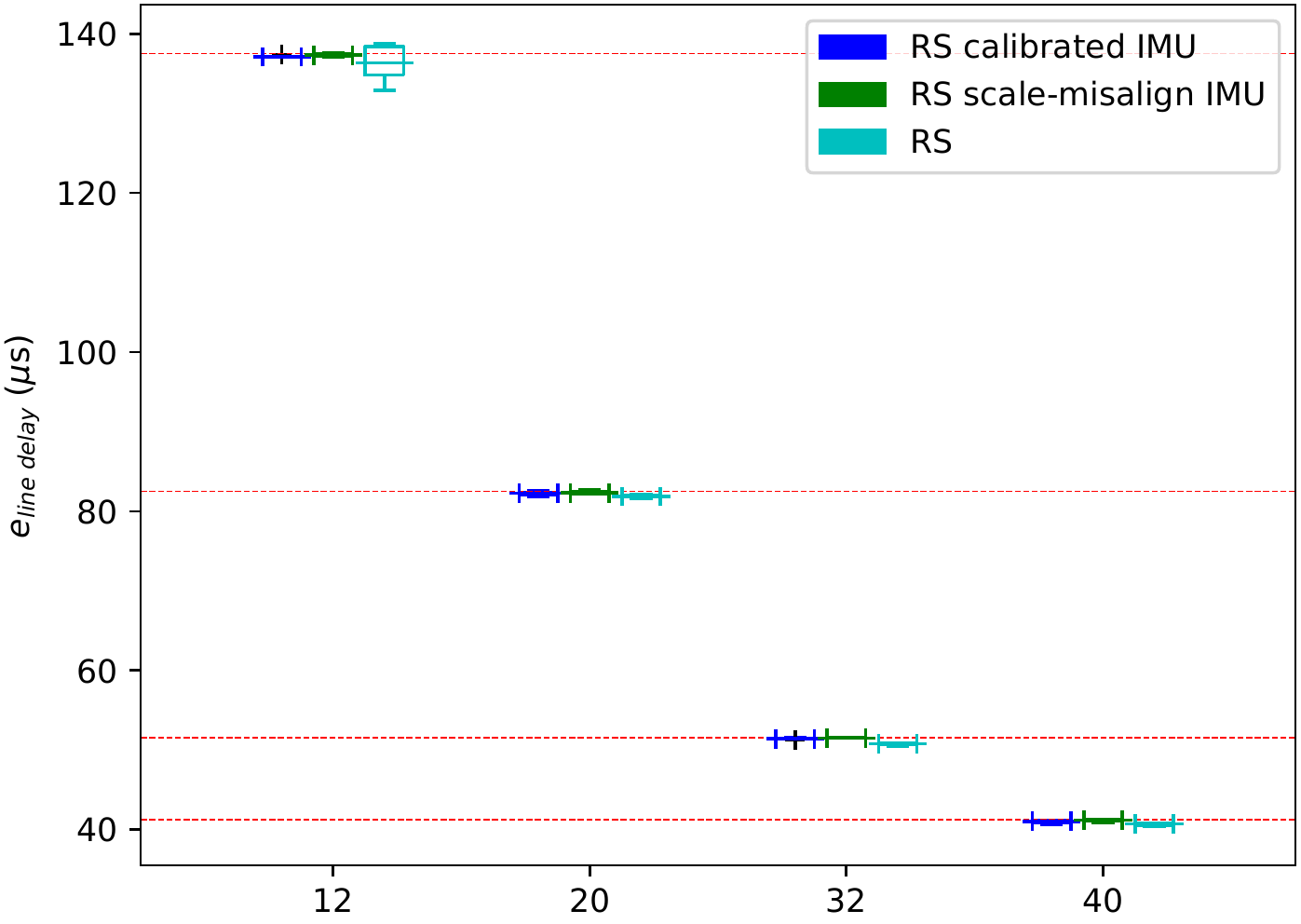}
\caption{
For the Bluefox-Artemis data, line delays estimated by the proposed method with both the calibrated IMU model (RS calibrated IMU)
and the scale-misalignment IMU model (RS scale-misalign IMU), and by 
the RS camera calibration tool (denoted by RS) \cite{othRolling2013}.
The reference values computed from the Bluefox specs, 137.5, 82.5, 51.563, and 41.25 ms, are shown by dashed lines.}
\label{fig:bluefox-line-delay}
\end{figure}

\subsubsection{Bluefox-Artemis}
\label{subsubsec:bluefox}
For the Bluefox camera and Artemis IMU assembly, the intrinsic parameters were estimated with the aforementioned image sequence by the calibration tool in Kalibr.
Since the reference extrinsic parameters were unavailable,
the errors in $\mathbf{T}_{CI}$ were computed relative to the values measured by hand, $\tilde{\mathbf{T}}_{CI}$,
\begin{equation}
\tilde{\mathbf{T}}_{CI} = \begin{bmatrix}
0 & -1 & 0 & 0.000 \pm 0.003 \\
1 & 0 & 0 & 0.098 \pm 0.003 \\
0 & 0 & 1 & -0.023 \pm 0.003
\end{bmatrix}.
\end{equation}
The accurate reference line delay is the ratio of line length $N_l$ = 1650 provided in datasheet and the pixel clock $P$ for the Bluefox camera.

The Bluefox-IMU data were processed by the four calibration methods.
The errors in $\mathbf{T}_{CI}$ and median reprojection errors are illustrated in Fig.~\ref{fig:bluefox-pose-reproj}.
Though the reference extrinsic parameters are inaccurate, 
from Fig.~\ref{fig:bluefox-pose-reproj}(a) and (b), we see that the RS effect prevented the GS calibration methods from achieving coherent spatial estimation,
while the proposed method had much smaller variances in extrinsic parameters.
Fig.~\ref{fig:bluefox-pose-reproj}(c) shows that the GS calibration methods had much greater reprojection errors than the RS calibration methods, 
similarly to Fig.~\ref{fig:ueye-pose-reproj}(c) with the uEye data.

The line delay estimates for the Bluefox camera are visualized in Fig.~\ref{fig:bluefox-line-delay}.
By comparing to the reference values, 
we see that the proposed method can accurately estimate line delays.
Unsurprisingly, it has better accuracy than the RS camera calibration tool \cite{othRolling2013} 
probably because the latter uses priors on angular and linear acceleration rather than real IMU measurements to constrain device motion.

\section{Conclusions}
In summary, to our knowledge, we present the first continuous-time spatiotemporal calibration method for a RS camera-IMU system.
The method is also able to estimate the inter-line delay of a RS.
By numerous simulation and real data tests with accurate reference values, 
we showed that RS could degrade the extrinsic calibration results of a GS-based
method by a few degrees in orientation and a few centimeters in translation.
These tests also validated that our approach achieved accurate and consistent line delay estimation and extrinsic calibration.

The future work is to extend the presented method to other sensor modalities, \eg, lidar-camera systems.

{\small
\bibliographystyle{ieee_fullname}
\bibliography{ms}

\begin{thebibliography}{10}\itemsep=-1pt

\bibitem{IEEE2006}
{{IEEE}} standard specification format guide and test procedure for single-axis
  laser gyros.
\newblock {\em IEEE Std 647-2006 (Revision of IEEE Std 647-1995)}, pages 1--83,
  2006.

\bibitem{bapatRolling2018}
Akash Bapat, True Price, and Jan-Michael Frahm.
\newblock Rolling shutter and radial distortion are features for high frame
  rate multi-camera tracking.
\newblock In {\em 2018 {{IEEE}}/{{CVF Conference}} on {{Computer Vision}} and
  {{Pattern Recognition}} ({{CVPR}})}, pages 4824--4833, {Salt Lake City, UT,
  USA}, June 2018. {IEEE}.

\bibitem{bassoRobust2018}
Filippo Basso, Emanuele Menegatti, and Alberto Pretto.
\newblock Robust intrinsic and extrinsic calibration of {{RGB}}-{{D}} cameras.
\newblock {\em IEEE Transactions on Robotics}, 34(5):1315--1332, 2018.

\bibitem{daiRolling2016}
Yuchao Dai, Hongdong Li, and Laurent Kneip.
\newblock Rolling shutter camera relative pose: {{Generalized}} epipolar
  geometry.
\newblock In {\em 2016 {{IEEE Conference}} on {{Computer Vision}} and {{Pattern
  Recognition}} ({{CVPR}})}, pages 4132--4140, {Las Vegas, NV, USA}, June 2016.

\bibitem{furgaleContinuoustime2012}
Paul Furgale, Timothy~D. Barfoot, and Gabe Sibley.
\newblock Continuous-time batch estimation using temporal basis functions.
\newblock In {\em 2012 {{IEEE International Conference}} on {{Robotics}} and
  {{Automation}} ({{ICRA}})}, pages 2088--2095, {Saint Paul, MN, USA}, May
  2012.

\bibitem{hedborgRolling2012}
Johan Hedborg, Per-Erik Forss{\'e}n, Michael Felsberg, and Erik Ringaby.
\newblock Rolling shutter bundle adjustment.
\newblock In {\em 2012 {{IEEE Conference}} on {{Computer Vision}} and {{Pattern
  Recognition}} ({{CVPR}})}, pages 1434--1441, {Providence, RI, USA}, June
  2012. {IEEE}.

\bibitem{huaiMobileARSensor2019}
Jianzhu Huai, Yujia Zhang, and Alper Yilmaz.
\newblock The mobile {{AR}} sensor logger for {{Android}} and {{iOS}} devices.
\newblock In {\em {{IEEE SENSORS}}}, pages 1--4, {Montreal, Canada}, Oct. 2019.

\bibitem{imAccurate2019}
Sunghoon Im, Hyowon Ha, Gyeongmin Choe, Hae-Gon Jeon, Kyungdon Joo, and In~So
  Kweon.
\newblock Accurate {{3D}} reconstruction from small motion clip for rolling
  shutter cameras.
\newblock {\em IEEE Transactions on Pattern Analysis and Machine Intelligence},
  41(4):775--787, Apr. 2019.

\bibitem{kangAutomatic2020}
Jaehyeon Kang and Nakju~L. Doh.
\newblock Automatic targetless camera\textendash{{LIDAR}} calibration by
  aligning edge with {{Gaussian}} mixture model.
\newblock {\em Journal of Field Robotics}, 37(1):158--179, 2020.

\bibitem{kannalaGeneric2006}
Juho Kannala and Sami~S. Brandt.
\newblock A generic camera model and calibration method for conventional,
  wide-angle, and fish-eye lenses.
\newblock {\em IEEE Transactions on Pattern Analysis and Machine Intelligence},
  28(8):1335--1340, Aug. 2006.

\bibitem{kerlDense2015}
Christian Kerl, Jorg Stuckler, and Daniel Cremers.
\newblock Dense continuous-time tracking and mapping with rolling shutter
  {{RGB}}-{{D}} cameras.
\newblock In {\em Proceedings of the {{IEEE International Conference}} on
  {{Computer Vision}} ({{ICCV}})}, pages 2264--2272, {Santiago, Chile}, Dec.
  2015.

\bibitem{kimObject2020}
Namhoon Kim, Junsu Bae, Cheolhwan Kim, Soyeon Park, and Hong-Gyoo Sohn.
\newblock Object distance estimation using a single image taken from a moving
  rolling shutter camera.
\newblock {\em Sensors}, 20(14):3860, Jan. 2020.

\bibitem{leeCalibration2018}
Chang-Ryeol Lee, Ju Yoon, and Kuk-Jin Yoon.
\newblock Calibration and noise identification of a rolling shutter camera and
  a low-cost inertial measurement unit.
\newblock {\em Sensors}, 18(7):2345, July 2018.

\bibitem{liVisionaidedInertialNavigation2014}
Mingyang Li and Anastasios~I. Mourikis.
\newblock Vision-aided inertial navigation with rolling-shutter cameras.
\newblock {\em The International Journal of Robotics Research},
  33(11):1490--1507, 2014.

\bibitem{mayeSelfsupervised2013}
J{\'e}r{\^o}me Maye, Paul Furgale, and Roland Siegwart.
\newblock Self-supervised calibration for robotic systems.
\newblock In {\em 2013 {{IEEE Intelligent Vehicles Symposium}} ({{IV}})}, pages
  473--480, {Gold Coast, Queensland, Australia}, 2013. {IEEE}.

\bibitem{nowickiSpatiotemporal2020}
Micha{\l}~R. Nowicki.
\newblock Spatiotemporal calibration of camera and {{3D}} laser scanner.
\newblock {\em IEEE Robotics and Automation Letters}, 5(4):6451--6458, Oct.
  2020.

\bibitem{othRolling2013}
Luc Oth, Paul Furgale, Laurent Kneip, and Roland Siegwart.
\newblock Rolling shutter camera calibration.
\newblock In {\em {{IEEE Conf}}. on {{Computer Vision}} and {{Pattern
  Recognition}} ({{CVPR}})}, pages 1360--1367, {Portland, OR, USA}, June 2013.
  {IEEE}.

\bibitem{ovrenTrajectory2019}
Hannes Ovr{\'e}n and Per-Erik Forss{\'e}n.
\newblock Trajectory representation and landmark projection for continuous-time
  structure from motion.
\newblock {\em The International Journal of Robotics Research}, 38(6):686--701,
  May 2019.

\bibitem{parkSpatiotemporal2020}
Chanoh Park, Peyman Moghadam, Soohwan Kim, Sridha Sridharan, and Clinton
  Fookes.
\newblock Spatiotemporal camera-{{LiDAR}} calibration: {{A}} targetless and
  structureless approach.
\newblock {\em IEEE Robotics and Automation Letters}, 5(2):1556--1563, 2020.

\bibitem{patron-perezSplinebased2015}
Alonso {Patron-Perez}, Steven Lovegrove, and Gabe Sibley.
\newblock A spline-based trajectory representation for sensor fusion and
  rolling shutter cameras.
\newblock {\em International Journal of Computer Vision}, 113(3):208--219, July
  2015.

\bibitem{qinGeneral26}
Kaihuai Qin.
\newblock General matrix representations for {{B}}-splines.
\newblock In {\em Proceedings {{Pacific Graphics}} '98}, {Singapore}, 26.

\bibitem{rehderExtending2016}
Joern Rehder, Janosch Nikolic, Thomas Schneider, Timo Hinzmann, and Roland
  Siegwart.
\newblock Extending kalibr: {{Calibrating}} the extrinsics of multiple {{IMUs}}
  and of individual axes.
\newblock In {\em {{IEEE Intl}}. {{Conf}}. on {{Robotics}} and {{Automation}}
  ({{ICRA}})}, pages 4304--4311, {Stockholm, Sweden}, May 2016.

\bibitem{rehderGeneralApproachSpatiotemporal2016}
Joern Rehder, Roland Siegwart, and Paul Furgale.
\newblock A general approach to spatiotemporal calibration in multisensor
  systems.
\newblock {\em IEEE Transactions on Robotics}, 32(2):383--398, Apr. 2016.

\bibitem{ringabyEfficient2012}
Erik Ringaby and Per-Erik Forss{\'e}n.
\newblock Efficient video rectification and stabilisation for cell-phones.
\newblock {\em International Journal of Computer Vision}, 96(3):335--352, 2012.

\bibitem{saurerSparse2016}
Olivier Saurer, Marc Pollefeys, and Gim~Hee Lee.
\newblock Sparse to dense {{3D}} reconstruction from rolling shutter images.
\newblock In {\em 2016 {{IEEE Conference}} on {{Computer Vision}} and {{Pattern
  Recognition}} ({{CVPR}})}, pages 3337--3345, {Las Vegas, NV, USA}, June 2016.
  {IEEE}.

\bibitem{schubertRollingshutter2019}
David Schubert, Nikolaus Demmel, Lukas {von Stumberg}, Vladyslav Usenko, and
  Daniel Cremers.
\newblock Rolling-shutter modelling for direct visual-inertial odometry.
\newblock Technical report, {Technical University of Munich, Germany}, Nov.
  2019.

\bibitem{sommerEfficient2020}
Christiane Sommer, Vladyslav Usenko, David Schubert, Nikolaus Demmel, and
  Daniel Cremers.
\newblock Efficient derivative computation for cumulative {{B}}-splines on
  {{Lie}} groups.
\newblock In {\em 2020 {{IEEE}}/{{CVF Conference}} on {{Computer Vision}} and
  {{Pattern Recognition}} ({{CVPR}})}, pages 11145--11153, June 2020.

\bibitem{sunScalability2020}
Pei Sun, Henrik Kretzschmar, Xerxes Dotiwalla, Aurelien Chouard, Vijaysai
  Patnaik, Paul Tsui, James Guo, Yin Zhou, Yuning Chai, and Benjamin Caine.
\newblock Scalability in perception for autonomous driving: {{Waymo}} open
  dataset.
\newblock In {\em Proceedings of the {{IEEE}}/{{CVF Conference}} on {{Computer
  Vision}} and {{Pattern Recognition}} ({{CVPR}})}, pages 2446--2454, June
  2020.

\bibitem{usenkoDouble2018}
Vladyslav Usenko, Nikolaus Demmel, and Daniel Cremers.
\newblock The double sphere camera model.
\newblock In {\em 2018 {{International Conference}} on {{3D Vision}}
  ({{3DV}})}, pages 552--560, {Verona, Italy}, Sept. 2018. {IEEE}.

\bibitem{zhuangSurvey2018}
Yuan Zhuang, Luchi Hua, Longning Qi, Jun Yang, Pan Cao, Yue Cao, Yongpeng Wu,
  John Thompson, and Harald Haas.
\newblock A survey of positioning systems using visible {{LED}} lights.
\newblock {\em IEEE Communications Surveys Tutorials}, 20(3):1963--1988, 2018.

\end{thebibliography}
}

\twocolumn[\centering \section*{\LARGE \bf Supplementary Material \\ \enskip}]
This supplementary material presents Allan variance analysis results of
an OpenLog Artemis board,
and extrinsic calibration results of the uEye camera with data captured in global shutter (GS) mode.

\section*{A. Allan variance analysis}
\begin{figure}[h]
\centering
\includegraphics[width=0.9\linewidth]{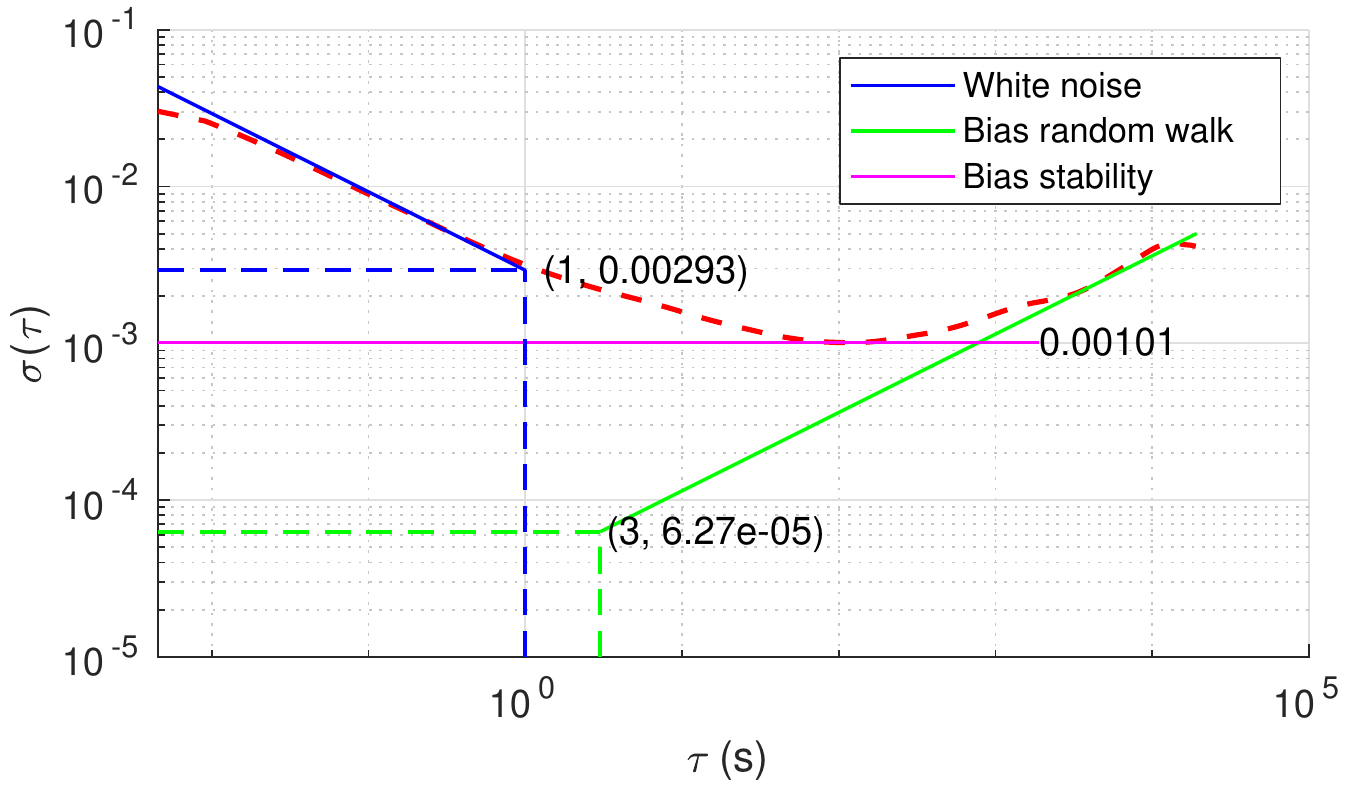} \\ (a) \\
\includegraphics[width=0.9\linewidth]{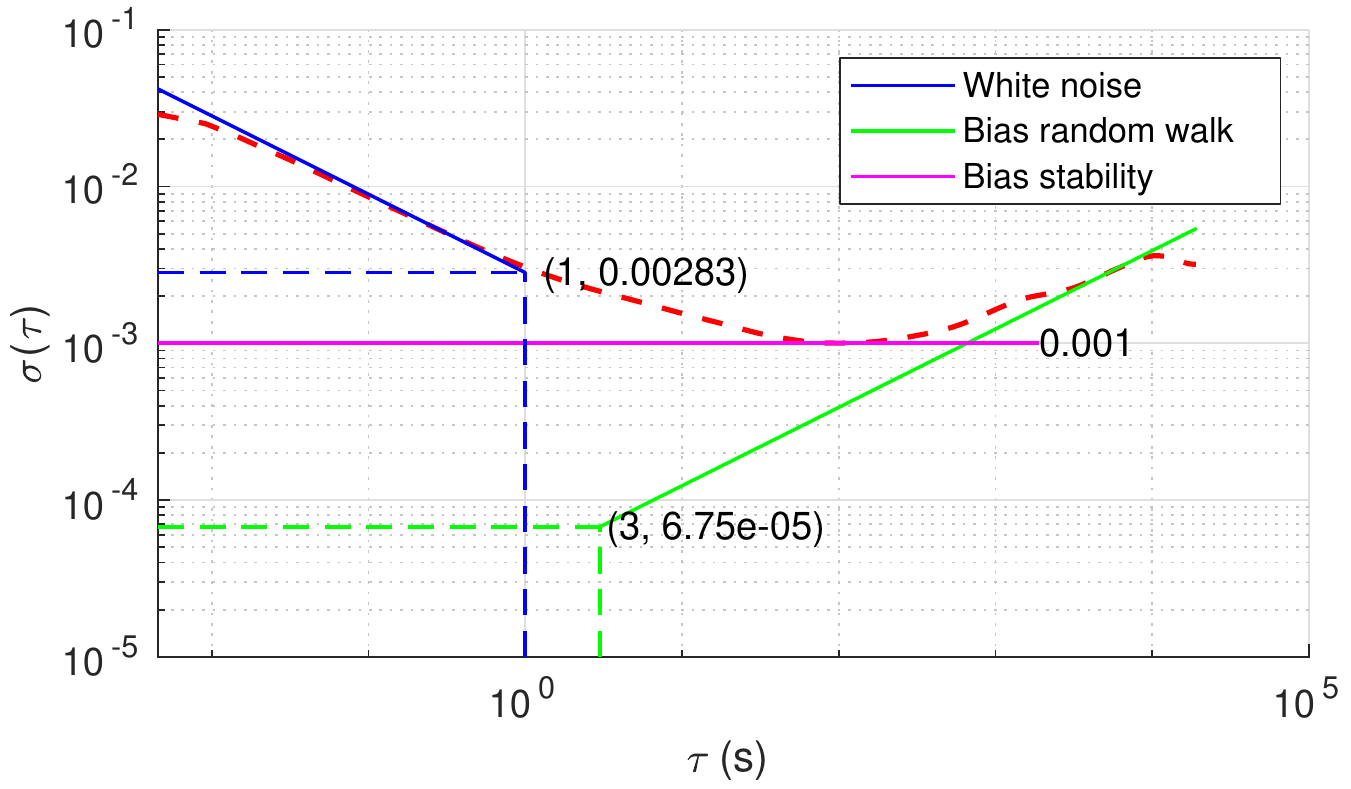} \\ (b) \\
\includegraphics[width=0.9\linewidth]{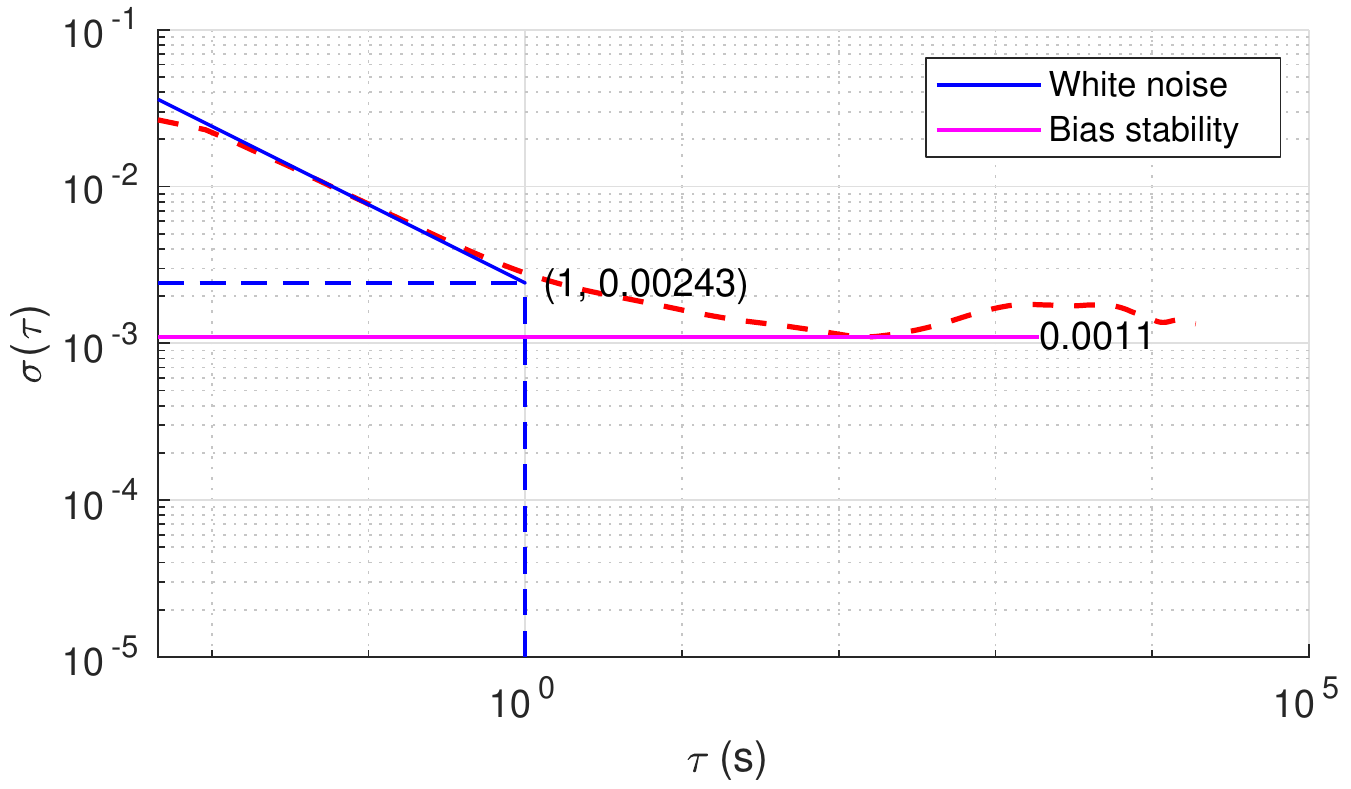} \\ (c)
\caption{Allan variance analysis results for the accelerometer on $x$(a), $y$(b), and $z$(c) axis of the Artemis board.}
\label{fig:artemis-allan-accel}
\end{figure}

\begin{figure}[h]
\centering
\includegraphics[width=0.9\linewidth]{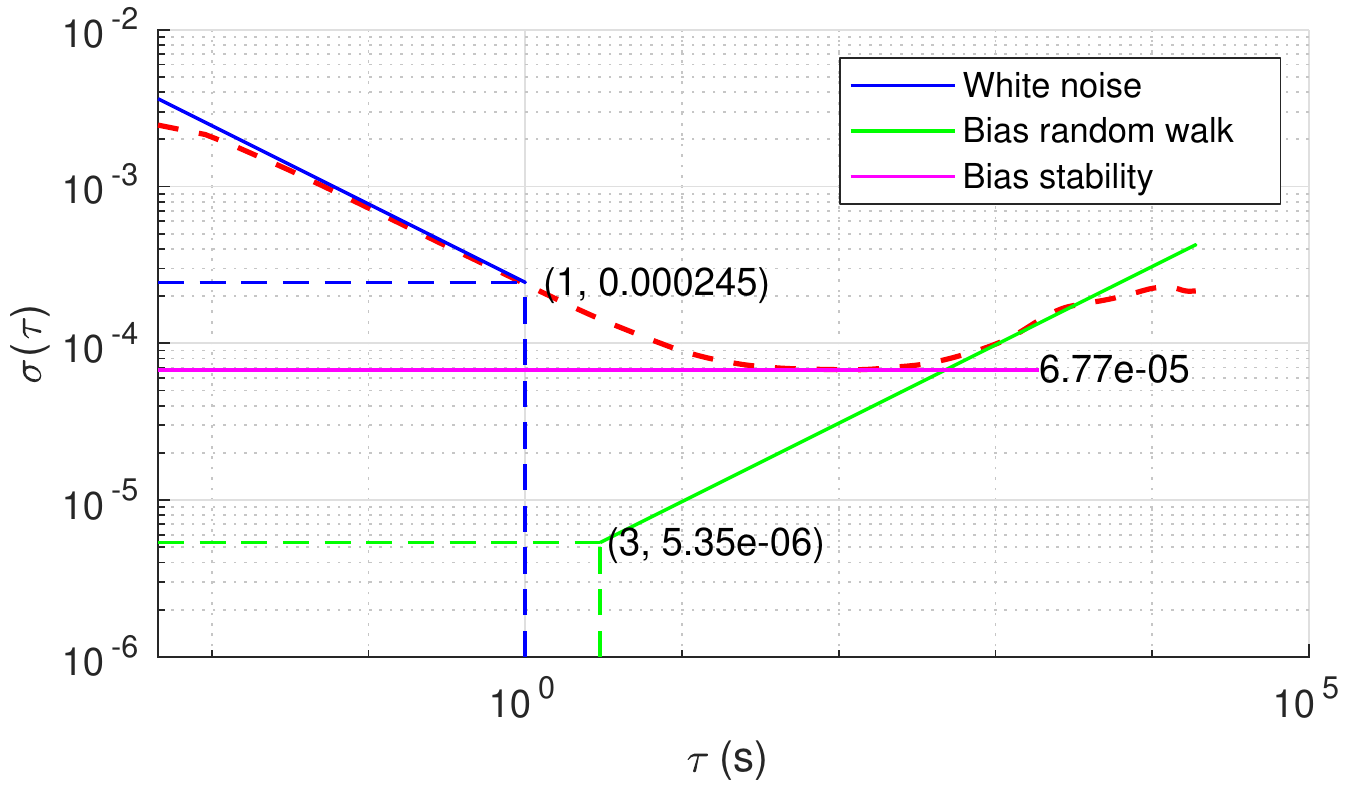} \\ (a) \\
\includegraphics[width=0.9\linewidth]{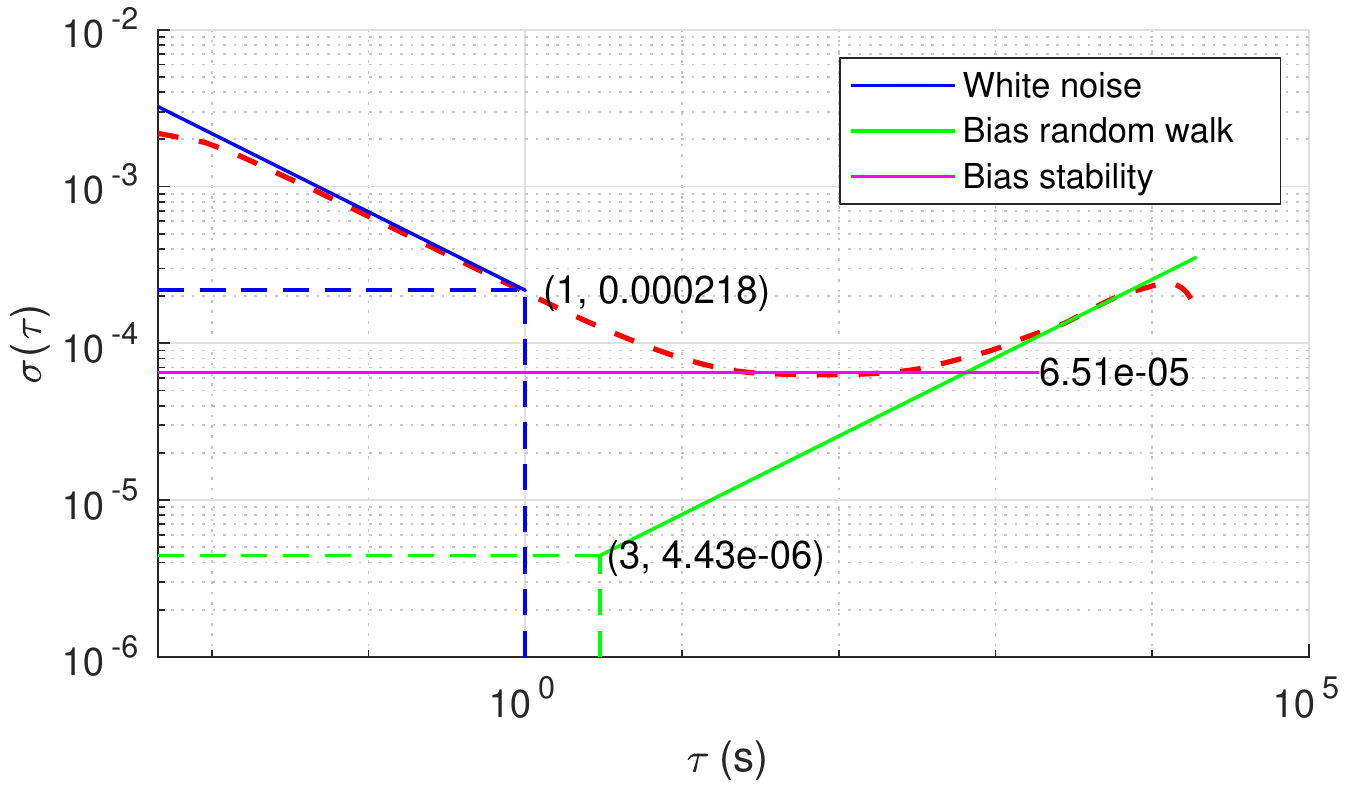} \\ (b) \\
\includegraphics[width=0.9\linewidth]{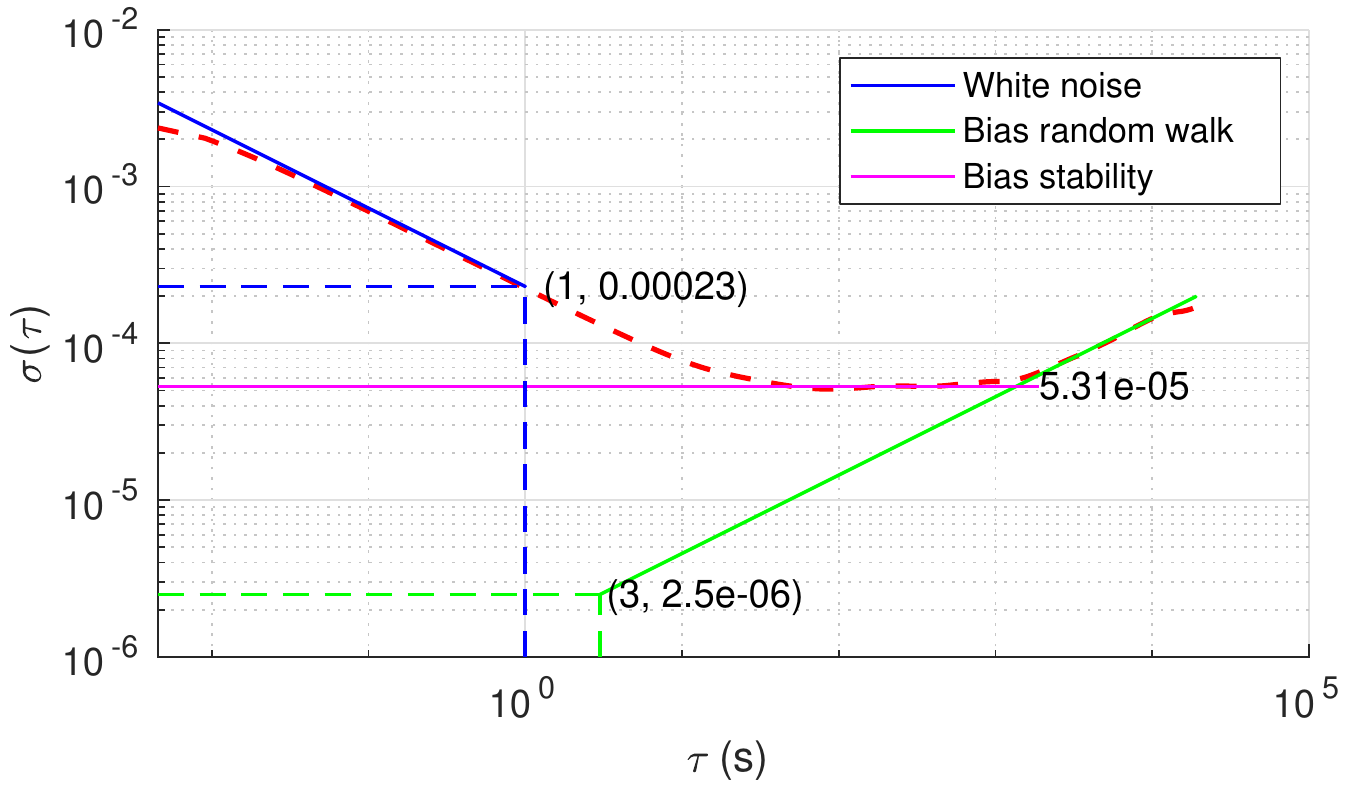} \\ (c)
\caption{Allan variance analysis results for the gyroscope on $x$(a), $y$(b), and $z$(c) axis of the Artemis board.}
\label{fig:artemis-allan-gyro}
\end{figure}

To characterize noises of the InvenSense ICM-20948 IMU on the OpenLog Artemis board,
the Allan variance analysis technique \cite{IEEE2006} was used to analyze its static data.
A 17-hour data at about 230Hz were captured from our lab at night while the Artemis board was placed on a table
with the $z$-axis of the board roughly along gravity.
The median of the temperatures recorded by the IMU was 27.36$^\circ$C, and its standard deviation was 1.36$^\circ$C.
The Allan standard deviations were computed for three accelerometers and three gyroscopes, 
shown in Fig.~\ref{fig:artemis-allan-accel} and Fig.~\ref{fig:artemis-allan-gyro}, respectively.
These figures are annotated with 
the interpreted values for the white noise, bias stability, and bias random walk,
whose corresponding slopes are -1/2, 0, and 1/2, respectively, 
according to the standard \cite{IEEE2006}.

The noise parameters and their averages are compiled in Table~\ref{tab:artemis-noise-values}.
For comparison, the reference noise density values read from the ICM-20948 datasheet are appended.
From Table~\ref{tab:artemis-noise-values}, we see that the noise densities from the Allan variance analysis are reasonably close to those from the datasheet.

\begin{table*}[h]
\centering
\caption{Noise parameters obtained from the Allan variance analysis for accelerometers and gyroscopes in Artemis.}
\label{tab:artemis-noise-values}
\begin{tabular}{c|ccc|ccc}
\hline
\multirow{2}{*}{Axis} & \multicolumn{3}{c|}{Accelerometer} & \multicolumn{3}{c}{Gyroscope} \\ \cline{2-7} 
 &
  \multicolumn{1}{l}{\begin{tabular}[c]{@{}l@{}}White noise\\ $m/(s^2 \sqrt{Hz})$\end{tabular}} &
  \multicolumn{1}{l}{\begin{tabular}[c]{@{}l@{}}Bias stability\\ $m/s^2$\end{tabular}} &
  \multicolumn{1}{l|}{\begin{tabular}[c]{@{}l@{}}Bias random walk\\ $m/(s^3 \sqrt{Hz})$\end{tabular}} &
  \multicolumn{1}{l}{\begin{tabular}[c]{@{}l@{}}White noise\\ $rad/(s \sqrt{Hz})$\end{tabular}} &
  \multicolumn{1}{l}{\begin{tabular}[c]{@{}l@{}}Bias stability\\ $rad/s$\end{tabular}} &
  \multicolumn{1}{l}{\begin{tabular}[c]{@{}l@{}}Rate random walk\\ $rad/(s^2 \sqrt{Hz})$\end{tabular}} \\ \hline
$x$                   & 2.93E-03   & 1.01E-03  & 6.27E-05  & 2.45E-04 & 6.77E-05 & 5.35E-06 \\ \hline
$y$                   & 2.83E-03   & 1.00E-03  & 6.75E-05  & 2.18E-04 & 6.51E-05 & 4.43E-06 \\ \hline
$z$                   & 2.43E-03   & 1.10E-03  & X         & 2.30E-04 & 5.31E-05 & 2.50E-06 \\ \hline
Mean                  & 2.73E-03   & 1.04E-03  & 6.51E-05  & 2.31E-04 & 6.20E-05 & 4.09E-06 \\ \hline
Datasheet & 2.26E-03 & X & X & 2.62E-04 & X & X \\ \hline
\end{tabular}
\end{table*}

\section*{B. uEye-Artemis calibration with GS data}
To see the variation of spatiotemporal calibration across different settings except the RS effect,
we carried out camera-IMU calibration with data captured by the uEye camera in GS mode.
These data were in fact captured along with the RS data for the test in Section~\ref{subsubsec:ueye}, 
including four sets of GS data captured at pixel clocks, 12, 20, 32, and 40 MHz.
Each set contained five one-minute sessions.

\begin{figure}[h]
\centering
\includegraphics[width=0.75\linewidth]{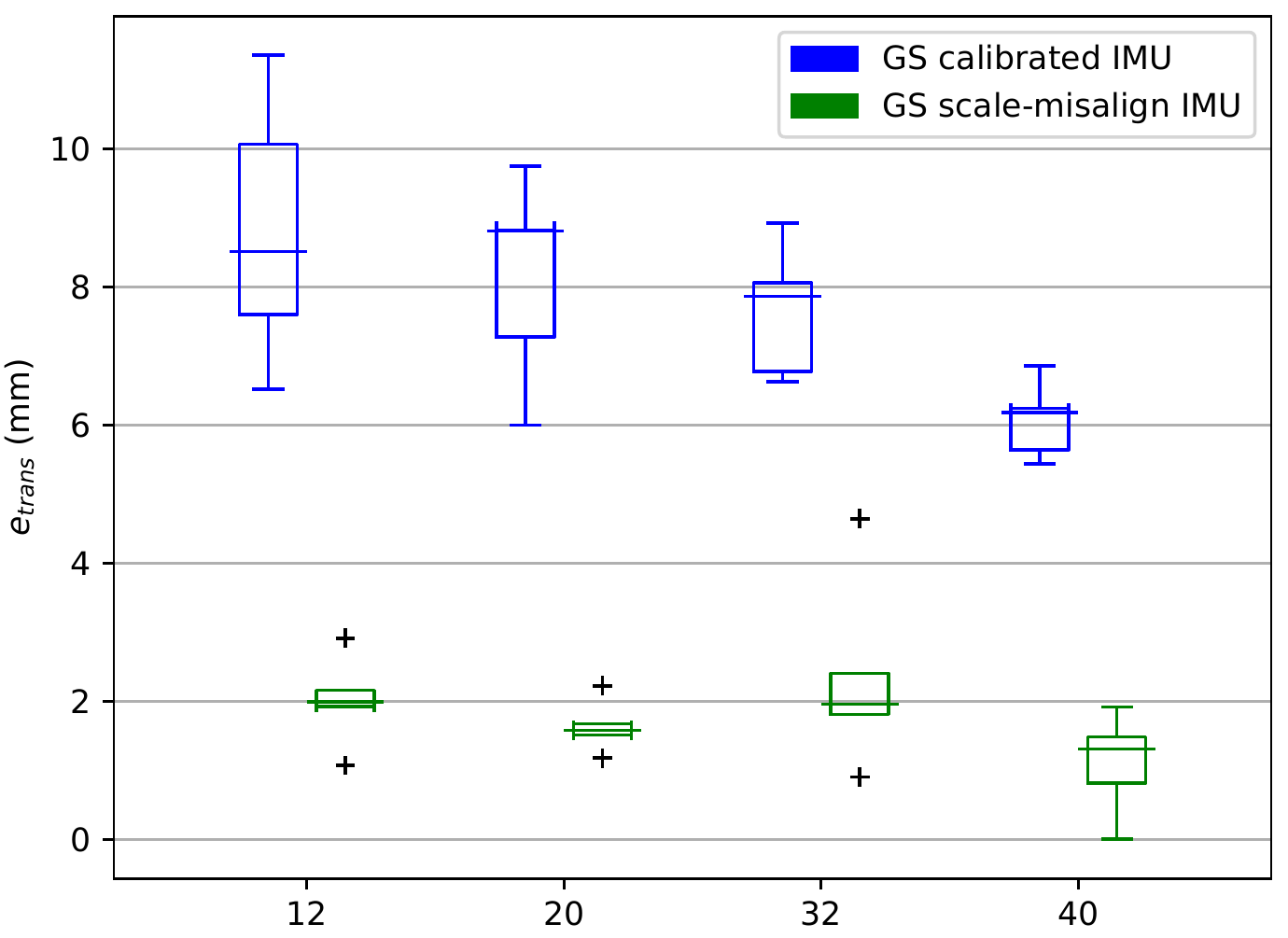} \\ (a) \\
\includegraphics[width=0.75\linewidth]{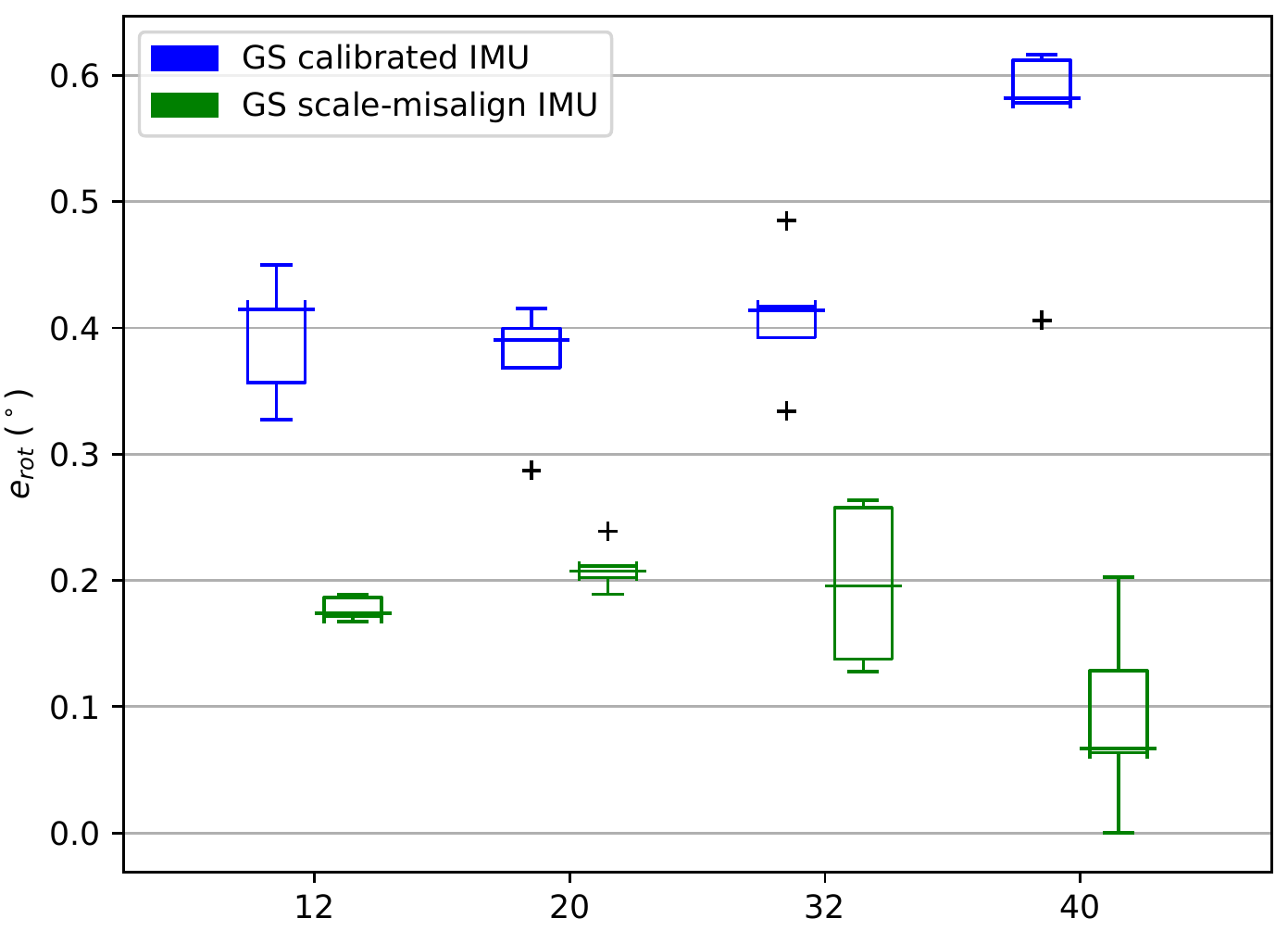} \\ (b) \\
\includegraphics[width=0.75\linewidth]{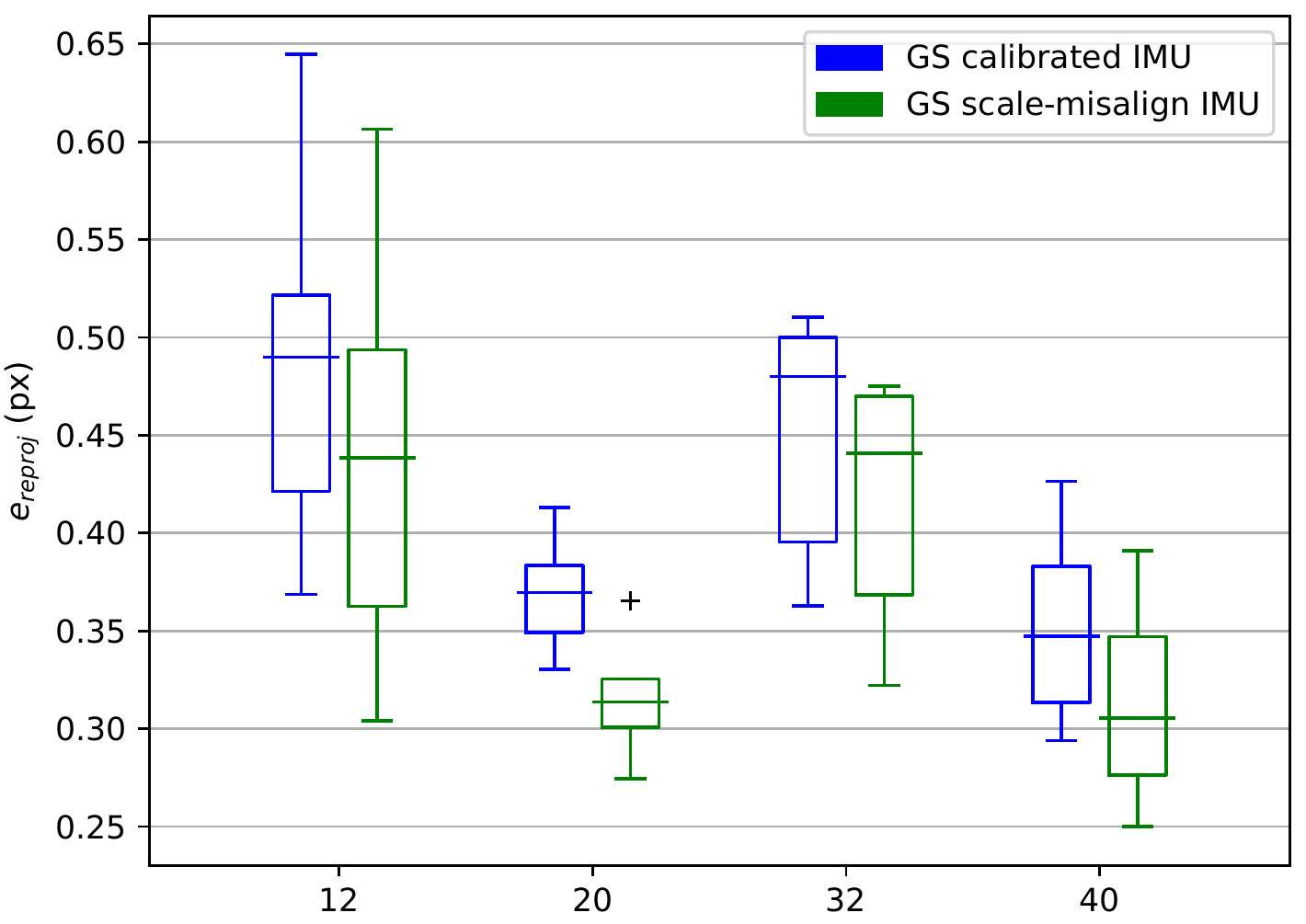} \\ (c)
\caption{Extrinsic calibration results of the uEye camera with data captured in global shutter (GS) mode at pixel clocks, 10, 20, 32, and 40 MHz.
Deviations in translation (a) and rotation (b) relative to the extrinsic
parameters estimated with a session of GS data at pixel clock 40 MHz.
(c) The median reprojection errors for calibration with both the calibrated and scale-misalignment IMU models.
}
\label{fig:ueye-artemis-gs}
\end{figure}

In calibrating the uEye-Artemis system, the same camera intrinsic parameters as in Section~\ref{subsubsec:ueye} were adopted.
These 20 sessions were processed with both the calibrated and scale-misalignment IMU models.
We computed the translation and rotation errors of the extrinsic parameters relative to the reference value 
$\mathbf{T}_{CI}$ in \eqref{eq:ueye_T_CI}, shown in Fig.~\ref{fig:ueye-artemis-gs}(a) and (b), respectively.
From these figures, we see that the estimated translations and rotations had typically small variances.
Especially with the GS scale-misalign IMU calibration method, the translation variations were mostly less than 2 mm, 
and the rotation variations mostly less than 0.2$^\circ$.
These variations arising from different settings in GS mode and different sessions were 
significantly smaller than the calibration deviations when ignoring the RS effect,
which were about 15 mm in translation and 0.4$^\circ$ in rotation even at pixel clock 40 MHz (see Fig.~\ref{fig:ueye-pose-reproj}(a-b)).

Boxplots of the median reprojection errors for the two calibration methods, GS calibrated IMU, and GS scale-misalign IMU, 
are shown in Fig.~\ref{fig:ueye-artemis-gs}(c), 
which shows that subpixel reprojection errors were attained when the GS data were processed by the GS camera calibration method.
This contrasted with the reprojection errors (often greater than 2 pixels) when
the RS data were processed by the GS camera calibration method (see Fig.~\ref{fig:ueye-pose-reproj}).

Overall, these additional results show that the RS effect is much more pronounced than the calibration uncertainty due to different sessions of data,
and the baseline GS camera-IMU calibration methods worked well for GS data.

\end{document}